\documentclass[a4paper,11pt]{article}
\usepackage{jheppub}

\pdfoutput=1
\usepackage{hyperref}
\usepackage{graphicx}
\usepackage[utf8]{inputenc}
\usepackage{xspace}
\usepackage{amsmath,amssymb}
\usepackage{bbm}
\usepackage{xcolor}
\usepackage{booktabs}
\usepackage{subcaption}
\usepackage{mathtools}
\usepackage[ruled,norelsize]{algorithm2e}
\usepackage{empheq}
\usepackage{cancel}
\usepackage{multirow}
\usepackage{slashed}
\usepackage{bbold}
\usepackage{cjhebrew}
\usepackage{textcomp} 
\usepackage{wasysym}

\DeclareFontFamily{U}{rcjhbltx}{}
\DeclareFontShape{U}{rcjhbltx}{m}{n}{<->rcjhbltx}{}
\newcommand{\RSV}{RSV}

\DeclareMathOperator*{\SumInt}{%
\mathchoice%
  {\ooalign{$\displaystyle\sum$\cr\hidewidth$\displaystyle\int$\hidewidth\cr}}
  {\ooalign{\raisebox{.14\height}{\scalebox{.7}{$\textstyle\sum$}}\cr\hidewidth$\textstyle\int$\hidewidth\cr}}
  {\ooalign{\raisebox{.2\height}{\scalebox{.6}{$\scriptstyle\sum$}}\cr$\scriptstyle\int$\cr}}
  {\ooalign{\raisebox{.2\height}{\scalebox{.6}{$\scriptstyle\sum$}}\cr$\scriptstyle\int$\cr}}
}

\title{Three-loop rapidity anomalous dimension for jet-veto cross sections}

\author[a,b,c]{Samuel Abreu,}
\author[d,e]{Jonathan R.~Gaunt,}
\author[a]{Pier Francesco Monni,}
\author[f]{Luca Rottoli,}
\author[g]{\\Robert Szafron}

\affiliation[a]{CERN, Theoretical Physics Department, CH-1211 Geneva
  23, Switzerland}
\affiliation[b]{Higgs Centre for Theoretical Physics, School of Physics and Astronomy,
 The University of Edinburgh, Edinburgh EH9 3FD, Scotland, United Kingdom}
\affiliation[c]{
Departamento de F\'isica Te\'orica y del Cosmos, Universidad de Granada, Campus de Fuentenueva, E-18071 Granada, Spain}
\affiliation[d]{Department of
  Physics and Astronomy, University of Manchester, Manchester M13 9PL,
  United Kingdom}
\affiliation[e]{Department of Physics, University of Cyprus, Nicosia 1678, Cyprus}
\affiliation[f]{Dipartimento di Fisica G. Occhialini, Universit\`a degli Studi di Milano-Bicocca and INFN,
Sezione di Milano-Bicocca, Piazza della Scienza 3,20126 Milano, Italy}
\affiliation[g]{Department of Physics, Brookhaven National Laboratory,
Upton, N.Y., 11973, U.S.A.}

\emailAdd{samuel.abreu@ugr.es}
\emailAdd{gaunt.jonathan@ucy.ac.cy}
\emailAdd{pier.monni@cern.ch}
\emailAdd{luca.rottoli@unimib.it}
\emailAdd{rszafron@bnl.gov}

\preprint{CERN-TH-2026-165}
\abstract{
    We present the first calculation of the third-order corrections to the rapidity anomalous dimension for jet-veto cross sections.
    This is a crucial ingredient for the resummation of these scattering observables in key processes such as colour-singlet and heavy-quark-pair production at hadron colliders.
    More precisely, our results enable the next-to-next-to-next-to-leading logarithmic (N$^3$LL) resummation of the jet-veto cross section for colour singlet production at the LHC, with immediate phenomenological relevance.
    The anomalous dimension is extracted from a calculation of the three-loop soft function in soft-collinear effective field theory, and is presented as a function of the jet radius $R$ for the clustering algorithms of the generalised-$k_t$ family.
    To handle the complexity associated with the clustering algorithm, we formulate a semi-numerical computational approach.
    This strategy is itself a result of the paper, and can be extended to the calculation of other anomalous dimensions and soft functions at three loops.
    }

\begin{document}
\setlength{\parskip}{0pt}
\maketitle
\flushbottom

\section{Introduction}
\label{sec:intro}

Jet vetoes are a common feature in a broad class of physics studies at the Large Hadron Collider (LHC). A tight cut on jet transverse momenta serves either to suppress background or to classify signal events into categories defined by jet multiplicities. 
From a theoretical viewpoint, the precise description of such jet-veto cross sections requires the all-order resummation of towers of logarithms of the ratio of the jet veto to the hard scale of the scattering process. In the simplest case of the colour-singlet production at hadron colliders, this problem has received considerable attention in past literature, with the next-to-next-to-leading logarithmic (NNLL) calculation currently defining the state of the art~\cite{Banfi:2012jm,Becher:2013xia,Stewart:2013faa,Banfi:2013eda,Banfi:2015pju}.

In a series of recent articles~\cite{Abreu:2022sdc,Abreu:2022zgo}, we have been systematically calculating the perturbative ingredients necessary to increase the accuracy of jet-veto cross section to N$^3$LL. In this article, we present the calculation of the last missing piece.
We consider the production of a generic colour-singlet system of squared invariant mass~$Q^2$, with a veto on
the transverse momentum of the leading jet, $p_{T}^{\rm jet} < p_{T}^{\rm veto}$, defined in the generalised-$k_t$ family of clustering algorithms~\cite{Catani:1993hr,Dokshitzer:1997in,Wobisch:1998wt,Cacciari:2008gp}. In the limit
$p_{T}^{\rm veto}\ll Q$, the cross section, differential in the kinematics of the colour-singlet system, is described by the following factorisation theorem obtained in the framework of soft-collinear effective field theory (SCET)~\cite{Bauer:2000yr,Bauer:2001yt,Bauer:2002nz,Beneke:2002ph,Beneke:2002ni}~\footnote{At high loop orders, observables of the kind studied here are likely affected by coherence-violating corrections~\cite{Forshaw:2006fk,Forshaw:2008cq,Forshaw:2009fz,Banfi:2010xy,DuranDelgado:2011tp,Gaunt:2014ska,Zeng:2015iba,Forshaw:2021fxs,Becher:2021zkk,Becher:2023mtx,Boer:2024hzh,Banfi:2025mra,Dasgupta:2025cgl,Becher:2026kbr}. Such corrections enter beyond the leading-$N_c$ level and start at the five loop order for colour singlet production.}
\begin{align}
\label{eq:factorisation}
\frac{d\sigma(p_{T}^{\rm veto})}{d\Phi_{\rm Born}} &\equiv
\sum_{F=q,g}
|{ A}^F_{\rm Born}|^2\,{\cal H}^F(Q;\mu)\notag\\
  &\times \,{\cal B}^F_n(x_1,Q,p_{T}^{\rm veto},R^2;\mu,\nu) \,{\cal B}^F_{\bar n}(x_2,Q,
  p_{T}^{\rm veto},R^2;\mu,\nu)\,{\cal S}^F(p_{T}^{\rm veto},R^2;\mu,\nu)\,.
\end{align}
Here, $d\Phi_{\rm Born}$ denotes the full Born phase-space measure
including the flux factor, ${ A}^F_{\rm Born}$ is the Born
amplitude for the production of the colour-singlet system, and $\mu$
and $\nu$ denote the renormalisation and rapidity regularisation
scales, respectively. 
All quantities on the second line of Eq.~\eqref{eq:factorisation}
depend explicitly on the jet radius $R$.
The index $F$ indicates the flavour configuration of the initial
state, i.e.~either $q\bar{q}$ ($F=q$) or $gg$ ($F=g$), and for
simplicity, we will drop it when referring to the ingredients of the
factorisation theorem~\eqref{eq:factorisation}.

In Eq.~\eqref{eq:factorisation}, the hard function ${\cal H}$
describes the dynamics at large virtuality scales $\mu\sim Q$, and
therefore contains purely virtual contributions. 
For most colour-singlet processes of phenomenological relevance it is now known at two loops~\cite{Anastasiou:2002zn, Idilbi:2006dg, Ahrens:2009cxz, Gehrmann:2011ab,Gehrmann:2015ora}, while three-loop results exist for the simplest processes such as Higgs and Drell-Yan pair production~\cite{Baikov:2009bg,Gehrmann:2010ue}.
The two beam functions ${\cal B}_n$ and ${\cal B}_{\bar n}$ describe the
collinear dynamics of radiation along the light-cone directions
$n^\mu$ and ${\bar n}^{\mu}$. They have recently been computed at two loops in Refs.~\cite{Abreu:2022zgo,Bell:2022nrj, Bell:2024epn}.
Finally, the soft function ${\cal S}$ describes the dynamics of soft
radiation emitted off the two initial-state partons. Its complete two-loop computation is presented in Refs.~\cite{Abreu:2022sdc} (see also~\cite{Bell:2020yzz}).

Each of the above ingredients satisfies a system of evolution equations whose solution achieves the resummation of jet-veto logarithms to all perturbative orders.
Specifically, the hard, beam, and soft functions satisfy a Renormalisation Group
Equation (RGE) describing the evolution in the renormalisation scale $\mu$ (see
e.g.~Refs.~\cite{Becher:2012qa,Becher:2013xia,Stewart:2013faa,Tackmann:2012bt,Jaiswal:2014yba} and references therein). In addition, each beam function ${\cal B}_i$ for $i=n,\,\bar n$ and the soft function ${\cal S}$ satisfy two evolution equations in the rapidity regularization scale $\nu$, which read
\begin{align}
 \label{eq:beam_nu}
\frac{d}{d\ln\nu}\ln {\cal B}_i(x,Q,p_{T}^{\rm veto},R^2;\mu,\nu) =& \,\,\,\,\,\,\,2
\int_{p_{T}^{\rm veto}}^{\mu}\frac{d\mu^\prime}{\mu^\prime}\Gamma_{\rm
  cusp}(\alpha_s(\mu^\prime)) -\frac{1}{2} \gamma_\nu(p_{T}^{\rm veto},R^2)\,,\\
\label{eq:soft_nu} 
\frac{d}{d\ln\nu}\ln {\cal S}(p_{T}^{\rm veto},R^2;\mu,\nu) =&
-4\int_{p_{T}^{\rm veto}}^{\mu}\frac{d\mu^\prime}{\mu^\prime}\Gamma_{\rm
  cusp}(\alpha_s(\mu^\prime)) + \gamma_\nu(p_{T}^{\rm veto},R^2)\,.
\end{align}
As $\mathcal{B}_i$ and $\mathcal{S}$, the anomalous dimensions $\Gamma_{\rm cusp}$ and $\gamma_\nu$ have a dependence on the flavour configuration $F$ of the initial state, but we suppress this dependence for simplicity.
To achieve N$^3$LL accuracy, the known fixed-order results for $\mathcal{H}$, $\mathcal{B}_{i}$ and $\mathcal{S}$ must be supplemented with the cusp anomalous dimension at four loops~\cite{Henn:2019swt}, the soft, hard, and collinear anomalous dimensions at three loops~\cite{Moch:2004pa,Gehrmann:2010ue,Li:2014afw}, and, finally, the hitherto unknown rapidity anomalous dimensions $\gamma_\nu^F$ at three loops whose calculation is the main topic of this paper.
We note that the $p_{T}^{\rm veto}$ dependence in $\gamma_\nu(p_{T}^{\rm veto},R^2)$ is solely encoded in the strong coupling which is evaluated at the scale $p_{T}^{\rm veto}$, that is 
\begin{equation}\label{eq:gammanu-coefficients}
\gamma_\nu(p_{T}^{\rm veto},R^2) = \sum_{n=1} \frac{\alpha^n_s(p_{T}^{\rm veto})}{(4\pi)^n}\gamma_\nu^{(n)}(R^2)\,.
\end{equation}
The three-loop contributions $\gamma_\nu^{(3)}(R^2)$ is extracted from a three-loop calculation of the soft function ${\cal S}$ for colour-singlet production. We present our results as a function of the jet radius $R$ for the main jet algorithms belonging to the generalised-$k_t$ family~\cite{Cacciari:2008gp}.
They can be obtained from Ref.~\cite{monni_2026_22914110}.

The article is organized as follows. In Sec.~\ref{sec:definition} we introduce the jet-veto soft function and discuss the perturbative structure relevant for the extraction of the rapidity anomalous dimension. In Sec.~\ref{sec:computation} we present the computational strategy based on a decomposition into a simplified reference observable and a remainder. In Sec.~\ref{sec:gammaRSV} we determine the rapidity anomalous dimension associated with the reference observable by relating it to the transverse-momentum-dependent soft function. In Sec.~\ref{sec:deltagamma2} we illustrate our method with an explicit two-loop example. The full three-loop calculation of the jet-veto rapidity anomalous dimension is presented in Sec.~\ref{sec:gamma3}, and the final results together with a set of non-trivial consistency checks are collected in Sec.~\ref{sec:results}. Finally, Sec.~\ref{sec:conclusions} contains our conclusions. Additional technical details of the calculation are provided in Appendix~\ref{app:details}.

\section{Perturbative structure of the jet-veto soft function}
\label{sec:definition}\sloppy
To extract the rapidity anomalous dimension, we consider the soft function ${\cal S}$, defined as
\begin{equation}
\label{eq:soft_def}
{\cal S}(p_{T}^{\rm veto},R^2;\mu,\nu) = \frac{1}{d_F}{\rm Tr}\left\{\langle
  0 | Y_{n}^{\dagger} Y_{\bar n} \,{\mathbb M}(p_{T}^{\rm veto},R^2)\,
  Y_{\bar n}^{\dagger} Y_{n}| 0 \rangle\right \}\,,
\end{equation}
where the soft Wilson line $Y_n$ is defined in terms of the soft gauge
field $A^{(s)}_a$ as~\cite{Bauer:2001yt}
\begin{equation}
Y_{n}(x) = {\cal P}\exp\left\{i g_s\int_{-\infty}^0 ds\, n\cdot
    A^{(s)}_a(x+n s){\mathbb T}^a\right\}\,, 
\end{equation}
and analogously for $Y_{\bar n}$.
The colour-charge operator ${\mathbb T}^a$ is understood to be in the
adjoint (fundamental) representation of $SU(N_C)$ for gluon (quark)
initiated processes. The trace in Eq.~\eqref{eq:soft_def} is performed
over colour indices and the normalisation factor $d_F$ denotes the
dimension of the representation, i.e.~$d_F=N_C^2-1$ for
gluon-initiated reactions and $d_F=N_C$ for quark-initiated ones.
The operator ${\mathbb M}(p_{T}^{\rm veto},R^2)$ acts on a given state
of $X_s$ soft particles $|X_s\rangle$ by applying a veto
$p_{T}^{\rm veto}$ on the final-state jets of jet radius $R$, obtained
with a clustering 
algorithm within the generalised-$k_t$ family 
\cite{Catani:1993hr,Dokshitzer:1997in,Wobisch:1998wt,Cacciari:2008gp}.

To proceed with the calculation of the soft function, we introduce a complete set of soft states in
Eq.~\eqref{eq:soft_def} and recast it as
\begin{equation}
\label{eq:soft_def_states}
{\cal S}(p_{T}^{\rm veto},R^2;\mu,\nu) = \frac{1}{d_F}\SumInt_{X_s} {\rm Tr}\left\{{\cal M}(p_{T}^{\rm veto},R^2)\,\langle
  0 | Y_{n}^{\dagger} Y_{\bar n} |X_s\rangle 
\langle X_s| Y_{\bar n}^{\dagger} Y_{n}| 0 \rangle\right \}\,,
\end{equation}
where ${\cal M}(p_{T}^{\rm veto},R^2)$ is the measurement function associated with the operator ${\mathbb M}(p_{T}^{\rm veto},R^2)$, which
schematically reads
\begin{equation}
  \label{eq:soft_def_measure}
{\cal M}(p_{T}^{\rm veto},R^2) = \Theta(p_{T}^{\rm veto}-p_t^{j_1}(p_1,\dots,p_n))\,.
\end{equation}
We have introduced $p_t^{j_1}(p_1,\dots,p_n)$ which clusters the particles with momenta $p_1$ through $p_n$ in $X_s$ into jets and returns the magnitude of the transverse momentum of the leading jet, defined with respect to the beam direction. The clustering depends on the jet radius $R$ (we suppress this dependence in $p_t^{j_1}$ for brevity) and on the specific jet algorithm. Within the generalised-$k_t$ family, the algorithm is defined in terms of the distances

\begin{equation}\label{eq:gen-kt-measures}
d_{ij} = \min\{p_{ti}^{2 p},p_{tj}^{2p}\} \frac{(\Delta y_{i j})^2 +
(\Delta\phi_{i j})^2}{R^2}\,,\quad d_{iB}=p_{ti}^{2 p}\,,
\end{equation}
with $p=-1$ for the anti-$k_t$~\cite{Cacciari:2008gp}, $p=0$ for the
Cambridge-Aachen~\cite{Dokshitzer:1997in,Wobisch:1998wt}, and $p=1$
for the $k_t$~\cite{Catani:1993hr} algorithm.
Here $p_{ti}$ denotes the (magnitude of the) transverse momentum of particle $i$, and $\Delta y_{i j}$ and $\Delta\phi_{i
  j}$ are the relative rapidity and azimuthal angle between particles
$i$ and $j$, respectively.

The soft function~\eqref{eq:soft_def_states} contains both ultra-violet (UV) and rapidity divergences. In our calculation, the UV divergences are regulated using conventional dimensional regularization (in the $\overline{\rm MS}$ scheme). A direct consequence of this is that we need to define the clustering algorithm, specifically the distances in Eq.~\eqref{eq:gen-kt-measures}, in a way that is agnostic to the dimensionality of the transverse component of particle momenta. A suitable generalisation is obtained by computing the azimuthal difference as
\begin{equation}\label{eq:deltaPhiGen}
    \Delta\phi_{ij} = \arccos\frac{\vec{p}_{ti}\cdot\vec{p}_{tj}}{p_{ti}\,p_{tj}}\,,
\end{equation}
where now $\vec{p}_{t}$ is a $D-2$ dimensional euclidean vector.
We always use the folded azimuthal separation  between jets, so $\Delta\phi_{ij} \in [0,\pi]$. 
The rapidity divergences are regularised by means of the exponential regulator of Ref.~\cite{Li:2016axz}. This is defined by modifying the phase-space integration measure for each real emission, such that
\begin{equation}
  \label{eq:regulator_def}
\prod_{i} d^D p_i \delta(p^2_i)\theta(p^0_i)\to  \prod_{i} d^D p_i \delta(p^2_i)\theta(p^0_i)\exp\left[ -\frac{e^{-\gamma_E}}{\nu}(n \cdot p_i + \bar{n} \cdot p_i)\right]\,,
\end{equation}
where $\nu$ denotes the rapidity regularization scale featuring in Eq.~\eqref{eq:factorisation}. The regularised soft function is obtained by taking the Laurent expansion around $\nu \to +\infty$, after the phase-space integrals over $p_i$ have been evaluated. 
Its logarithmic dependence on $\nu$ is controlled by Eq.~\eqref{eq:soft_nu}, which we will use to determine the rapidity anomalous dimension $\gamma_\nu$. In fact, in this paper we will neglect terms that are $\nu$ independent as, while they contribute to the soft function, they are not required to determine $\gamma_\nu$.

The soft function admits a perturbative expansion in powers of
the strong coupling $\alpha_s$,
\begin{equation}
{\cal S}(p_{T}^{\rm veto},R^2;\mu,\nu) = 1 + \sum_{n=1}^3\frac{\alpha^n_s}{(4\pi)^n} {\cal S}^{(n)}(p_{T}^{\rm veto},R^2;\mu,\nu) + {\cal O}(\alpha_s^4)\,.
\end{equation}
In the regularization scheme adopted here, the two-loop ($n=2$) result  was presented in Ref.~\cite{Abreu:2022sdc}. The three-loop correction ${\cal S}^{(3)}$ receives contributions from tree-level triple-real, one-loop double-real, and two-loop single-real configurations. 
However, as will become clear in Sec.~\ref{sec:computation}, we will in fact not directly require the soft single-real two-loop amplitude (which was computed in Refs.~\cite{Badger:2004uk,Duhr:2013msa}).

We will denote tree-level \textit{squared amplitudes}  for the emission of $n$ soft partons by ${\cal A}^{(0)}(p_1,\dots,p_n) \equiv |A^{(0)}(p_1,\dots,p_n)|^2$, and the corresponding one-loop corrections by  ${\cal A}^{(1)}(p_1,\dots,p_n) \equiv 2\Re\left\{A^{(0)\,\dagger}(p_1,\dots,p_n)A^{(1)}(p_1,\dots,p_n)\right\}$. More precisely, these denote renormalised amplitudes, expanded in terms of the renormalised coupling $\alpha_s(\mu)$ in the $\overline{\rm MS}$ scheme, which is defined in terms of the bare coupling $\alpha_s$ through
\begin{equation}\label{eq:MSbar}
 \alpha_s \mu^{2\epsilon} = Z_{\alpha_s} \mu^{2\epsilon} \alpha_s(\mu)\,,\quad Z_{\alpha_s} = \frac{e^{\gamma_E\,\epsilon}}{(4\pi)^\epsilon}\left(1-\frac{b_0}{\epsilon}\frac{\alpha_s(\mu)}{4\pi} + \dots\right)\,,\quad b_0 = \frac{11}{3} C_A - \frac{4}{3}  T_R n_f\,.
\end{equation}
In particular, this implies that the one-loop squared amplitudes include terms proportional to $b_0$ related to the renormalisation procedure.

An important organizational principle in our calculation will be the separation between \emph{correlated} and \emph{independent} contributions, which has been exploited, e.g., in Refs.~\cite{Banfi:2004yd,Banfi:2014sua,Gangal:2016kuo,Bizon:2017rah,Banfi:2018mcq}. This separation can be made explicit at the level of the squared amplitudes.
For instance, the double-gluon squared amplitudes ${\cal A}^{(0)}_{gg}(p_1,p_2)$~\cite{Catani:1999ss} can be written as
\begin{equation}\label{eq:Atildegg}
     {\cal A}^{(0)}_{gg}(p_1,p_2)=\tilde{\cal A}^{(0)}_{gg}(p_1,p_2) + {\cal A}^{(0)}_g(p_1)\,{\cal A}^{(0)}_g(p_2)\,,
\end{equation}
where $\tilde{\cal A}^{(0)}_{gg}(p_i,p_j)$ denotes the correlated contribution and ${\cal A}^{(0)}_g(p_i)$ denotes the single-gluon squared amplitude~\cite{Bassetto:1983mvz}.
Similarly, the  triple-gluon squared amplitudes~\cite{Catani:2019nqv} can 
be written as
\begin{align}
{\cal A}^{(0)}_{ggg}(p_1,p_2,p_3)&=\tilde{\cal A}^{(0)}_{ggg}(p_1,p_2,p_3)  \notag\\
&+ \tilde{\cal A}^{(0)}_{gg}(p_1,p_2)\,{\cal A}^{(0)}_g(p_3) + \tilde{\cal A}^{(0)}_{gg}(p_1,p_3)\,{\cal A}^{(0)}_g(p_2) + \tilde{\cal A}^{(0)}_{gg}(p_2,p_3)\,{\cal A}^{(0)}_g(p_1)\notag\\
& + {\cal A}^{(0)}_g(p_1)\,{\cal A}^{(0)}_g(p_2)\,{\cal A}^{(0)}_g(p_3)\,.
\end{align}
The advantage of these decompositions is that the correlated terms $\tilde{\cal A}_{g \dots }(p_1,p_2,\dots)$ describe configurations where the particles $p_1,p_2,\dots$ have commensurate rapidities. As such, they vanish whenever one of these particles is far in rapidity from the others. This implies that the correlated term $\tilde{\cal A}_{g \dots }(p_1,p_2,\dots)$ only has a single rapidity divergence associated with the absolute rapidity of the whole correlated $p_1,p_2,\dots$ system becoming large. This divergence can be mapped onto the integration over the rapidity of only one of the $p_1,p_2,\dots$ emissions, which renders rapidity regularization in Eq.~\eqref{eq:regulator_def} straightforward.
The soft squared amplitude for the emission of a quark pair and a gluon \cite{Catani:2022hkb} can be decomposed in a similar way,
reading
\begin{align}
{\cal A}^{(0)}_{gq\bar{q}}(p_1,p^{q}_2,p^{\bar q}_3) &= \tilde{\cal A}^{(0)}_{q\bar{q}}(p^{q}_2,p^{\bar q}_3)\,{\cal A}^{(0)}_g(p_1) +\tilde{\cal A}^{(0)}_{gq\bar{q}}(p_1,p^{q}_2,p^{\bar q}_3)\,,
\end{align}
where~\cite{Catani:1999ss}
\begin{equation}\label{eq:Atildeqq}
   \tilde{\cal A}^{(0)}_{q\bar{q}}(p_i,p_j) = {\cal A}^{(0)}_{q\bar{q}}(p_i,p_j) \,.
\end{equation}
Finally, the soft one-loop double-real squared amplitudes can be obtained from the amplitudes computed in Refs.~\cite{Czakon:2022dwk}, and expressed as
\begin{align}
   {\cal A}^{(1)}_{gg}(p_1,p_2) &= {\cal A}^{(0)}_g(p_1)\,{\cal A}^{(1)}_g(p_2)+{\cal A}^{(1)}_g(p_1)\,{\cal A}^{(0)}_g(p_2)+\tilde{\cal A}^{(1)}_{gg}(p_1,p_2)\,,\\
   {\cal A}^{(1)}_{q\bar{q}}(p_1,p_2) &= \tilde{\cal A}^{(1)}_{q\bar{q}}(p_1,p_2)\,,
\end{align}
where the decomposition into correlated building blocks obeys the same organizing principle as for the tree-level amplitudes.

The above amplitudes are to be integrated over the dimensionally- and rapidity-regularised real-emission phase space, weighted by the measurement function~\eqref{eq:soft_def_measure}. An important technical comment concerns the explicit parametrization of Eq.~\eqref{eq:soft_def_measure} in $D=4-2\epsilon$ space-time dimensions. With up to two real emissions, one can always perform a Lorentz rotation on the $D$-dimensional momenta $p_i^\mu$ such that they lie in the $4$-dimensional subspace and their extra-dimensional components vanish.
The corresponding directions can then be trivially integrated
over. 
This is not possible, in general, with three or more emissions (hence in the triple-real corrections), where the $(D-4)$-dimensional components give non-trivial contributions. A direct consequence of this fact is that to evaluate the corresponding phase-space integrals we need to analytically continue the measurement function~\eqref{eq:soft_def_measure} beyond four dimensions (including the clustering algorithm, as discussed around Eq.~\eqref{eq:deltaPhiGen}). A suitable parametrization of the kinematics valid for any clustering algorithm of the generalised-$k_t$ family is given in Appendix~\ref{app:details}.

\subsection{Computational strategy and definition of the RSV soft function}
\label{sec:computation}\sloppy
In this section, we formulate our strategy to computing the soft function and extracting the rapidity anomalous dimension.
Our approach relies on decomposing the bare-level soft function as
\begin{equation}~\label{eq:decomposition}
{\cal S}(p_{T}^{\rm veto},R^2;\mu,\nu) = {\cal S}_{\rm \RSV}(p_{T}^{\rm veto};\mu,\nu) + \Delta {\cal S}(p_{T}^{\rm veto},R^2;\mu,\nu)\,.
\end{equation}
The quantity ${\cal S}_{\rm \RSV}$ denotes the soft function for a reference observable that we dub \textit{Reference Simplified Variable} in the following, and $\Delta {\cal S}$ denotes the difference between the jet-veto soft function and ${\cal S}_{\rm \RSV}$.
At variance with a similar decomposition used in Refs.~\cite{Abreu:2022sdc,Abreu:2022zgo}, here we note that the RSV is not a physical observable, but rather an infrared-safe quantity that we design to simplify its calculation as much as possible, as detailed in the rest of this section.
Our computational strategy involves an analytic determination of ${\cal S}_{\rm \RSV}$ and a numerical calculation of $\Delta {\cal S}(p_{T}^{\rm veto},R^2;\mu,\nu)$.

The definition of ${\cal S}_{\rm \RSV}$ builds on the decomposition of the soft squared amplitudes into independent and correlated contributions defined in the previous section. We consider the total sum $p_{12\dots}$ of momenta entering each correlated term $\tilde{\cal A}(p_1,p_2,\dots)$, which we refer to as \textit{correlated clusters}. 
For the simplest case in which the cluster involves a single emission, i.e.~${\cal A}_g(p)$, we consider $p^\mu$ as the cluster momentum. The \RSV{} measurement function is then defined by integrating each correlated term $\tilde{\cal A}(p_1,p_2,\dots)$ with a measurement function given by $\Theta(p_{T}^{\rm veto}-p_t^{\rm clust})$, where $p_t^{\rm clust}$ denotes the transverse component of the cluster's momentum w.r.t.~the beam direction. The measurement function defining ${\cal S}_{\rm \RSV}$ is thus
\begin{equation}
  \label{eq:RSV_measure}
{\cal M}_{\rm \RSV}(p_{T}^{\rm veto}) = \!\!\prod_{ i}\Theta(p_{T}^{\rm veto}-p_t^{\rm clust_i})=\Theta(p_{T}^{\rm veto}-\max\{p_t^{\rm clust_i}\})\,.
\end{equation}
This definition leads to the factorisation of phase-space integrals over multiple clusters into the product of the integrals of the individual clusters. This simplification is instrumental and allows us to determine ${\cal S}_{\rm \RSV}$ analytically. From its definition~\eqref{eq:RSV_measure}, it is clear that ${\cal S}_{\rm \RSV}$ is independent of the jet radius $R$.

The second term in the r.h.s.~of Eq.~\eqref{eq:decomposition}, $\Delta {\cal S}(p_{T}^{\rm veto},R^2;\mu,\nu)$, encodes the difference between the jet-veto soft function and the \RSV{} counterpart. The definition of the \RSV{} is such that $\Delta {\cal S}(p_{T}^{\rm veto},R^2;\mu,\nu)$ vanishes for final states with a single real emission, which explains why in our calculation we do not require the soft single-real two-loop amplitude. This implies that the singularity structure of the difference $\Delta {\cal S}(p_{T}^{\rm veto},R^2;\mu,\nu)$ is significantly simplified. In particular, at two loops $\Delta {\cal S}(p_{T}^{\rm veto},R^2;\mu,\nu)$ has the singularities typically appearing in a tree-level calculation and, at three-loops, it has the singularities typically appearing in a one-loop calculation. This drop by two perturbative orders of the singularity structure of $\Delta {\cal S}(p_{T}^{\rm veto},R^2;\mu,\nu)$ persists at higher loops. 

To set up the calculation of ${\cal S}_{\rm \RSV}$ and $\Delta {\cal S}(p_{T}^{\rm veto},R^2;\mu,\nu)$, we first discuss the renormalisation of the soft function ${\cal S}(p_{T}^{\rm veto},R^2;\mu,\nu)$. We recast the bare-level decomposition in Eq.~\eqref{eq:decomposition} as
\begin{equation}~\label{eq:decomposition-ren}
Z\,{\cal S}^{\rm R}(p_{T}^{\rm veto},R^2;\mu,\nu) = Z_{\rm \RSV}\,{\cal S}^{\rm R}_{\rm \RSV}(p_{T}^{\rm veto};\mu,\nu) + \Delta {\cal S}(p_{T}^{\rm veto},R^2;\mu,\nu)\,.
\end{equation}
Here, we used the fact that both ${\cal S}$ and ${\cal S}_{\rm \RSV}$ renormalise locally in momentum space, and we denoted the corresponding renormalisation constants by $Z$ and $Z_{\rm \RSV}$, respectively. In this case, due to rapidity regularization, each renormalisation constant must be observable independent, which necessarily implies that $Z_{\rm \RSV} = Z$, as discussed in Refs.~\cite{Bauer:2020npd,Abreu:2022zgo}.
Expanding Eq.~\eqref{eq:decomposition-ren} in a perturbative series in the renormalised coupling $\alpha_s(\mu)$ leads to the following expressions for the expansion of $\Delta {\cal S}$ in terms of renormalised quantities
\begin{align}
\Delta {\cal S}^{(1)}(p_{T}^{\rm veto},R^2;\mu,\nu) &= 0\,,\label{eq:DS-1L}\\
\Delta {\cal S}^{(2)}(p_{T}^{\rm veto},R^2;\mu,\nu) &= {\cal S}^{(2)}(p_{T}^{\rm veto},R^2;\mu,\nu)-{\cal S}_{\rm \RSV}^{(2)}(p_{T}^{\rm veto};\mu,\nu) \notag\\&= {\cal S}^{(2),\,{\rm R}}(p_{T}^{\rm veto},R^2;\mu,\nu)-{\cal S}_{\rm \RSV}^{(2),\,{\rm R}}(p_{T}^{\rm veto};\mu,\nu)\,,\label{eq:DS-2L}\\
\Delta {\cal S}^{(3)}(p_{T}^{\rm veto},R^2;\mu,\nu) &= {\cal S}^{(3)}(p_{T}^{\rm veto},R^2;\mu,\nu)-{\cal S}_{\rm \RSV}^{(3)}(p_{T}^{\rm veto};\mu,\nu) \notag\\&= {\cal S}^{(3),\,{\rm R}}(p_{T}^{\rm veto},R^2;\mu,\nu)-{\cal S}_{\rm \RSV}^{(3),\,{\rm R}}(p_{T}^{\rm veto};\mu,\nu)\,\notag\\
&-Z^{(1)}\,\Delta {\cal S}^{(2)}(p_{T}^{\rm veto},R^2;\mu,\nu)\,,\label{eq:DS-3L}
\end{align}
where we used the fact that 
${\cal S}^{(1)}(p_{T}^{\rm veto},R^2;\mu,\nu)={\cal S}_{\rm \RSV}^{(1)}(p_{T}^{\rm veto};\mu,\nu)$ and
\begin{align}
 Z &= 1 - \sum_{n=1} \frac{\alpha_s^n(\mu)}{(4\pi)^n} Z^{(n)}\,,\\
 \Delta {\cal S}(p_{T}^{\rm veto},R^2;\mu,\nu) &= \sum_{n=1}\frac{\alpha_s^n(\mu)}{(4\pi)^n}\,\Delta {\cal S}^{(n)}(p_{T}^{\rm veto},R^2;\mu,\nu) \,.
\end{align}
We then see that while $\Delta {\cal S}^{(2)}$ is free of UV singularities, the three-loop correction $\Delta {\cal S}^{(3)}$ inherits the UV poles in the dimensional regulator from the one-loop counterterm $Z^{(1)}$.

We now explore the implication of this decomposition on the rapidity anomalous dimension. We consider the logarithmic derivative $\nu \frac{d}{d\nu}$ of the logarithm of the soft function, $\ln{\cal S}^{\rm R}(p_{T}^{\rm veto},R^2;\mu,\nu)$. Analogously to the decomposition~\eqref{eq:decomposition}, we can express this as the sum of the logarithmic derivative of $\ln{\cal S}_{\rm RSV}^{\rm R}(p_{T}^{\rm veto};\mu,\nu)$ and a correction term $\Delta\gamma_{\nu}(R^2)$.  
The former reads (see Eq.~\eqref{eq:soft_nu})
\begin{equation}
\label{eq:soft_nu_RSV} 
\frac{d}{d\ln\nu}\ln {\cal S}^{\rm R}_{\rm \RSV}(p_{T}^{\rm veto};\mu,\nu) =
-4\int_{p_{T}^{\rm veto}}^{\mu}\frac{d\mu^\prime}{\mu^\prime}\Gamma_{\rm
  cusp}(\alpha_s(\mu^\prime)) + \gamma_{\nu,\,{\rm \RSV}}(p_{T}^{\rm veto})\,,
\end{equation}
and the latter can be expressed as 
\begin{equation}
\Delta\gamma_{\nu}(p_{T}^{\rm veto},R^2)=\frac{d}{d\ln\nu}\ln \frac{{\cal S}^{\rm R}(p_{T}^{\rm veto},R^2;\mu,\nu)}{{\cal S}^{\rm R}_{\rm \RSV}(p_{T}^{\rm veto};\mu,\nu)}\equiv \frac{d}{d\ln\nu} \Sigma(p_{T}^{\rm veto},R^2;\mu,\nu)\,,
\end{equation}
where, to simplify the notation in the following sections,
we introduced
\begin{equation}
    \Sigma(p_{T}^{\rm veto},R^2;\mu,\nu)=\sum_{n=1}\frac{\alpha_s(\mu)^n}{(4\pi)^n}\Sigma^{(n)}\,,
\end{equation}
and we omit the dependence of $\Sigma^{(n)}$ on $p_{T}^{\rm veto}$, $R^2$, $\mu$, $\nu$.
Up to three loops, we have
\begin{align}
    \Sigma^{(1)}&= 0\,,\\
    \label{eq:sigmaExp}
    \Sigma^{(2)}&= \Delta {\cal S}^{(2)}(p_{T}^{\rm veto},R^2;\mu,\nu) = {\cal S}^{(2)}(p_{T}^{\rm veto},R^2;\mu,\nu)-{\cal S}_{\rm \RSV}^{(2)}(p_{T}^{\rm veto};\mu,\nu) \,,\\
    \Sigma^{(3)}&=  \Delta {\cal S}^{(3)}(p_{T}^{\rm veto},R^2;\mu,\nu)-\Delta {\cal S}^{(2)}(p_{T}^{\rm veto},R^2;\mu,\nu)\,{\cal S}_{\rm \RSV}^{(1)}(p_{T}^{\rm veto};\mu,\nu) \notag\\
    & = {\cal S}^{(3),\,R}(p_{T}^{\rm veto},R^2;\mu,\nu)-{\cal S}_{\rm \RSV}^{(3),\,R}(p_{T}^{\rm veto},R^2;\mu,\nu)\notag\\&-\,{\cal S}_{\rm \RSV}^{(1),\,R}(p_{T}^{\rm veto};\mu,\nu)\,\Delta {\cal S}^{(2)}(p_{T}^{\rm veto},R^2;\mu,\nu)\,.
\end{align}

From Eq.~\eqref{eq:soft_nu_RSV} we observe that the term proportional to $\Gamma_{\rm cusp}$ in the rapidity evolution equation is identical to the one in Eq.~\eqref{eq:soft_nu}. This implies that the quantity $\Delta\gamma_{\nu}(p_{T}^{\rm veto},R^2)$ contributes only to $\gamma_{\nu}(p_{T}^{\rm veto},R^2)$ which can be expressed as 
\begin{equation}\label{eq:RAD}
    \gamma_{\nu}(p_{T}^{\rm veto},R^2) = \gamma_{\nu,\,{\rm \RSV}}(p_{T}^{\rm veto}) + \Delta\gamma_{\nu}(p_{T}^{\rm veto},R^2)\,,
\end{equation}
where, in the convention of Eq.~\eqref{eq:gammanu-coefficients}
\begin{equation}\label{eq:delta-gammanu-coefficients}
\Delta\gamma_\nu(p_{T}^{\rm veto},R^2) = \sum_{n=1} \frac{\alpha^n_s(p_{T}^{\rm veto})}{(4\pi)^n}\Delta\gamma_\nu^{(n)}(R^2)\,.
\end{equation}
with
\begin{align}
\Delta\gamma_{\nu}^{(i)}(R^2) &= \frac{d \Sigma^{(i)}}{d\ln\nu}\label{eq:Dgamma}\,.
\end{align}
In Sec.~\ref{sec:gamma3} we will derive an integral representation for $\Delta\gamma_{\nu}^{(3)}(R^2)$. Combined with the three-loop expression for $\gamma_{\nu,\,{\rm \RSV}}(p_{T}^{\rm veto})$, this will allow us to determine the three-loop jet-veto rapidity anomalous dimension.

We close this discussion by noting that in order
to determine $\Delta\gamma_{\nu}^{(i)}(R^2)$ from 
$\Sigma^{(i)}$ we simply need to retain the $\ln\nu$
dependence in the latter. This further simplifies 
the handling of the rapidity regulator, since we
can neglect terms of order $\mathcal{O}(\nu^0)$ arising from 
the expansion of Eq.~\eqref{eq:regulator_def} after the 
$p_i$ integration has been performed.

\subsection{Bootstrapping $\gamma^\nu_{\rm \RSV}$ at three loops}
\label{sec:gammaRSV}
The first step in the calculation of the rapidity anomalous dimension~\eqref{eq:RAD} is the derivation of $\gamma_{\nu,\,{\rm \RSV}}(p_{T}^{\rm veto})$. This anomalous dimension governs the rapidity evolution of the RSV soft function, defined as in Eq.~\eqref{eq:soft_def_states} but with the measurement replaced by Eq.~\eqref{eq:RSV_measure}. As we will show in this section, ${\cal S}_{\rm \RSV}(p_{T}^{\rm veto};\mu,\nu)$ can be related to the transverse-momentum-dependent (TMD) soft function ${\cal S}_\perp(b^2;\mu,\nu)$. 
This relation will allow us to determine 
$\gamma_{\nu,\,{\rm \RSV}}(p_{T}^{\rm veto})$
from the results of Refs.~\cite{Li:2016ctv}, where ${\cal S}_\perp(b^2;\mu,\nu)$ was computed to three loops.

The link between the TMD and RSV soft functions relies on two key observations. The first is that, by construction, they both satisfy the non-abelian exponentiation theorem~\cite{Gatheral:1983cz,Frenkel:1984pz}. We can then write the two bare soft functions as\footnote{We express them in terms of the renormalised $\overline{\rm MS}$ coupling constant.}
\begin{align}
\begin{split}\label{eq:non-abelian-exponentiation-theorem}
    {\cal S}_\perp(b^2;\mu,\nu) &= \exp\left\{\sum_{n=1}\frac{\alpha^n_s(\mu)}{(4\pi)^n} s_\perp^{(n)}(b^2;\mu,\nu)\right\}\,,\\ 
    {\cal S}_{\rm \RSV}(p_{T}^{\rm veto};\mu,\nu) &= \exp\left\{\sum_{n=1}\frac{\alpha^n_s(\mu)}{(4\pi)^n} s_{\rm RSV}^{(n)}(p_{T}^{\rm veto};\mu,\nu)\right\}\,.
\end{split}
\end{align}

The $n$-th order term in the exponent has a simple representation in terms of the correlated clusters defined in Sec.~\ref{sec:definition}, as both the TMD and RSV soft functions are only sensitive to the total momentum of the radiation in each correlated cluster at a given perturbative order. We can schematically write 
\begin{align}
\begin{split}
    s_\perp^{(n)}(b^2;\mu,\nu) &= \sum_{j=0}^{n-1}\int d \Phi_\perp^{(n-j)} \tilde{\cal A}^{(j)}(p_1,\dots,p_{n-j})\,,\\
    s_{\rm RSV}^{(n)}(p_{T}^{\rm veto};\mu,\nu) &= \sum_{j=0}^{n-1}\int d \Phi_{\rm RSV}^{(n-j)} \tilde{\cal A}^{(j)}(p_1,\dots,p_{n-j})\,,
\end{split}
\end{align}
where we have defined
\begin{equation}
    d\Phi_\perp^{(n)}\equiv \prod_{k=1}^n[d p_k]\, e^{i \vec{b}\cdot\vec{p}_{tk}}\,,\quad d\Phi_{\rm RSV}^{(n)}\equiv \prod_{k=1}^n[d p_k] \,\Theta\left(p_{T}^{\rm veto}-|\sum_{k}^n \vec{p}_{tk}|\right)\,,
\end{equation}
and
\begin{equation}\label{eq:LIPS}
    [d p] \equiv \frac{d^D p}{(2\pi)^{D-1}} \,\delta(p^2)\,\theta(p^0)\,\exp\left[ -\frac{e^{-\gamma_E}}{\nu}(n \cdot p + \bar{n} \cdot p)\right]\,.
\end{equation}
We recall that, beyond tree level, each $\tilde{\cal A}^{(j\,>\,0)}(p_1,\dots,p_{n-j})$ also contains terms arising from the coupling renormalisation proportional to lower-order contributions $\tilde{\cal A}^{(0\leq i\,<\,j)}(p_1,\dots,p_{n-j})$.
From the above equations we see that in order to connect $s_\perp^{(n)}$ to $s_{\rm RSV}^{(n)}$ one simply needs to connect $d\Phi_\perp^{(n)}$ to $d\Phi_{\rm RSV}^{(n)}$. This is easily done by observing that in $D=4-2\epsilon$ dimensions
\begin{equation}\label{eq:dPhi-connection}
    d\Phi_{\rm RSV}^{(n)} = \frac{2^{\epsilon}\,p_{T}^{\rm veto}}{\Gamma(1-\epsilon)}\int_{0}^{+\infty} \!\!\!\!\!d b \,(b\,p_{T}^{\rm veto})^{-\epsilon} J_{1-\epsilon}(b \,p_{T}^{\rm veto})\,d\Phi_\perp^{(n)}\equiv {\cal G}^{(\epsilon)}_{p_{T}^{\rm veto}}[d\Phi_\perp^{(n)}]\,,
\end{equation}
where $J_n(x)$ denotes a Bessel function of the first kind. From Eqs.~\eqref{eq:non-abelian-exponentiation-theorem}--\eqref{eq:dPhi-connection} it follows that
\begin{equation}\label{eq:Srsv}
\ln {\cal S}_{\rm \RSV}(p_{T}^{\rm veto};\mu,\nu) = {\cal G}^{(\epsilon)}_{p_{T}^{\rm veto}}[\ln {\cal S}_\perp(b^2;\mu,\nu)]\,.
\end{equation}
We have verified this relation up to three loops with an explicit calculation of the difference ${\cal S}_{\rm \RSV}-{\cal G}^{(\epsilon)}_{p_{T}^{\rm veto}}[{\cal S}_\perp]$. 

From Eq.~\eqref{eq:Srsv} it is then straightforward to relate $\gamma_{\nu,\,{\rm \RSV}}$ to the rapidity anomalous dimension of the TMD soft function $\gamma_{\nu,\perp}$ at any loop order, defined in terms of the renormalised TMD soft function as
\begin{equation}
\label{eq:soft_nu_perp} 
\frac{d}{d\ln\nu}\ln {\cal S}^{\rm R}_{\perp}(b^2;\mu,\nu) =
-4\int_{\frac{b_0}{\sqrt{b^2}}}^{\mu}\frac{d\mu^\prime}{\mu^\prime}\Gamma_{\rm
  cusp}(\alpha_s(\mu^\prime)) + \gamma_{\nu,\,\perp}\left(\frac{b_0}{\sqrt{b^2}}\right)\,,\quad b_0 = 2\,e^{-\gamma_E}\,.
\end{equation}
Using the above equation, we find 
\begin{align}\label{eq:RAD_RSV}
\gamma_{\nu,\,{\rm \RSV}}(p_{T}^{\rm veto}) &= \gamma_{\nu,\perp}(p_{T}^{\rm veto}) - 2 \,{\cal G}^{(\epsilon = 0)}_{p_{T}^{\rm veto}}\left[\int_{\alpha_s(b_0/\sqrt{b^2})}^{\alpha_s(p_{T}^{\rm veto})}\frac{d\alpha_s}{\beta(\alpha_s)} \Gamma_{\rm cusp}(\alpha_s)\right] \notag\\&= \gamma_{\nu,\perp}(p_{T}^{\rm veto}) + \left(\frac{\alpha_s(p_{T}^{\rm veto})}{4\pi}\right)^3 \frac{32}{3} C_R b_0^2 \,\zeta_3 + {\cal O}(\alpha_s^4)\,,
\end{align}
where $\beta(\alpha_s)= d\alpha_s/d\ln\mu^2$ and $\gamma_{\nu,\perp}$ can be extracted from Refs.~\cite{Li:2016ctv,Vladimirov:2016dll}.
It is interesting to observe that due to the properties of the non-abelian exponentiation, $\gamma_{\nu,\perp}$ and $\gamma_{\nu,\,{\rm \RSV}}$ only differ by terms proportional to the QCD $\beta$ function.
As a side observation, we note that the same class of terms proportional to the $\beta$ function appearing in Eq.~\eqref{eq:RAD_RSV} also appear in the formulation of transverse-momentum resummation in momentum space (cf.~Section 2.4, and specifically Eq.~(2.70) and related discussion in Ref.~\cite{Bizon:2017rah}), where they originate from the transverse-momentum measure in direct space as opposed to its impact-parameter space counterpart.

\section{An example: $\Delta \gamma^\nu$ at two loops}
\label{sec:deltagamma2}
We now address the calculation of the correction term $\Delta\gamma_{\nu}(p_{T}^{\rm veto},R^2)$ defined in Eq.~\eqref{eq:RAD}.
To illustrate the procedure we start with the two-loop calculation, deferring the three-loop case to the next section. 
The full two-loop soft function in the rapidity regularization scheme adopted here was computed analytically in Ref.~\cite{Abreu:2022sdc}. 

To follow the procedure outlined in the previous section, we start from Eq.~\eqref{eq:Dgamma} and take the $\nu \frac{d}{d\nu}$ derivative of the quantity $\Sigma^{(2)}=\Delta {\cal S}^{(2)}$, see Eqs.~\eqref{eq:DS-2L} and \eqref{eq:sigmaExp}. We can now derive an integral representation of the difference between the two bare two-loop soft functions. We start by noticing that this difference has support only in the double-real configurations given that $\Delta {\cal S}^{(1)}(p_{T}^{\rm veto},R^2)=0$. We can then use the decomposition into the correlated (CC) and independent (II) contributions of the soft gluon squared amplitude given in Eq.~\eqref{eq:Atildegg}, together with the corresponding soft quark squared amplitude, which is purely correlated,  to recast $\Sigma^{(2)}$ as the sum of the following two terms
\begin{equation}\label{eq:S2_decomposition}
    \Sigma^{(2)} =  \Sigma^{(2)}_{\rm CC} + \Sigma^{(2)}_{\rm II} \,.
\end{equation}
Correspondingly, we can decompose the two-loop correction to $\Delta\gamma^{(2)}_{\nu}(R^2)$ as
\begin{equation}
    \Delta\gamma^{(2)}_{\nu}(R^2) = \Delta\gamma^{(2)}_{\nu,\,{\rm CC}}(R^2)+\Delta\gamma^{(2)}_{\nu,\,{\rm II}}(R^2) \,,
\end{equation}
as the $\nu \frac{d}{d\nu}$ derivatives of each term in Eq.~\eqref{eq:S2_decomposition}.

We have defined the contribution from two (identical) independent emissions as
\begin{equation}
    \Sigma^{(2)}_{\rm II}  = \frac{1}{2!}\int [d p_1] [d p_2] \,{\cal A}^{(0)}_g(p_1)\,{\cal A}^{(0)}_g(p_2) \left(\Theta(p_{T}^{\rm veto} - p_t^{j_1}(p_1,p_2))-\Theta(p_{T}^{\rm veto} - \max\{p_{t1}, p_{t2}\})\right)\,,\label{eq:S2II}
\end{equation}
where the measures ${\cal M}$ and ${\cal M}_{\rm \RSV}$ are defined in Eqs.~\eqref{eq:soft_def_measure},~\eqref{eq:RSV_measure}, respectively.
The second term in Eq.~\eqref{eq:S2_decomposition} encodes the contribution from two correlated emissions, and is defined as
\begin{align}
    \Sigma^{(2)}_{\rm CC}  = \int [d p_1] [d p_2] \, &\left(\Theta(p_{T}^{\rm veto} - p_t^{j_1}(p_1,p_2))-\Theta(p_{T}^{\rm veto} - |\vec{p}_{t1} + \vec{p}_{t2}|)\right) \notag\\&\times\left(\tilde{\cal A}^{(0)}_{q\bar{q}}(p_1,p_2)+\frac{1}{2!}\tilde{\cal A}^{(0)}_{gg}(p_1,p_2) \right)\,,\label{eq:S2CC}
\end{align}
where we include a combinatorial factor of $1/2!$ for the two indistinguishable gluon emissions, and $\tilde{\cal A}^{(0)}_{gg}$, $\tilde{\cal A}^{(0)}_{q\bar{q}}$ are defined in Eqs.~\eqref{eq:Atildegg},~\eqref{eq:Atildeqq}.
The jet veto partitions the phase space according to $\Theta(p_{T}^{\rm veto} - p_t^{j_1}(p_1,p_2)) = \Theta(\min(d_{1B},d_{2B}) - d_{12}) \Theta(p_{T}^{\rm veto} - |\vec{p}_{t1} + \vec{p}_{t2}|)+
 \Theta(  d_{12}-\min(d_{1B},d_{2B})) \Theta(p_{T}^{\rm veto} - \max\{p_{t1}, p_{t2}\})$.

In general, the integrals entering $\Sigma^{(i)}$ at a given loop order contain both IR (regulated in dimensional regularization) and rapidity divergences (regulated by the exponential regulator in Eq.~\eqref{eq:regulator_def}).
At two loops, a first simplification in the calculation of the integrals~\eqref{eq:S2II},~\eqref{eq:S2CC} is that they are free of $\epsilon$ poles, and therefore we can directly work in $D=4$ dimensions.
To handle the exponential rapidity regulator embedded in the phase-space measures $[d p_i]$ defined in Eq.~\eqref{eq:LIPS}, we consider the expression of the measures in Eqs.~\eqref{eq:S2II},~\eqref{eq:S2CC}:
\begin{align}
\Theta(p_{T}^{\rm veto} &- p_t^{j_1}(p_1,p_2))-\Theta(p_{T}^{\rm veto} - \max\{p_{t1}, p_{t2}\})\label{eq:2L-measure-II}\\&=\Theta(\min(d_{1B},d_{2B}) - d_{12}) \left(\Theta(p_{T}^{\rm veto} - |\vec{p}_{t1} + \vec{p}_{t2}|) - \Theta(p_{T}^{\rm veto} - \max\{p_{t1}, p_{t2}\})\right)\,,\notag\\
 \Theta(p_{T}^{\rm veto} &- p_t^{j_1}(p_1,p_2))-\Theta(p_{T}^{\rm veto} - |\vec{p}_{t1} + \vec{p}_{t2}|)\label{eq:2L-measure-CC}\\&=\Theta( d_{12}-\min(d_{1B},d_{2B})) \left(\Theta(p_{T}^{\rm veto} - \max\{p_{t1}, p_{t2}\}) - \Theta(p_{T}^{\rm veto} - |\vec{p}_{t1} + \vec{p}_{t2}|)\right)\,,\notag
%checked by RS and SA 28.08.2025
\end{align}
where we used the distances given in Eq.~\eqref{eq:gen-kt-measures}. At two loops, the clustering condition in Eqs.~\eqref{eq:2L-measure-II},~\eqref{eq:2L-measure-CC} is independent of the jet algorithm adopted (within the generalised-$k_t$ family) and reads
\begin{equation}
\Theta(\min(d_{1B},d_{2B}) - d_{12}) = \Theta(R^2 - (\Delta y_{1 2})^2 -
(\Delta\phi_{1 2})^2)\,.
\end{equation}
Owing to the choice we made for the RSV observable, the measurements in Eqs.~\eqref{eq:2L-measure-II},~\eqref{eq:2L-measure-CC} have support in configurations where the emissions are close in rapidity. Therefore, as we will see below, this implies that we can simplify the structure of the rapidity regularization by regulating only a single rapidity integral in the phase-space measure. This is expected: $\Delta {\cal S}^{(2)}$ must be linear in $\ln \nu$ since it directly contributes to $\Delta\gamma_{\nu}$ in Eq.~\eqref{eq:Dgamma}, and hence contain a single rapidity divergence.
As we will further clarify in the three-loop case, this property holds more generally at all loop orders, and it allows us to simplify considerably the treatment of the rapidity regulator in the calculation of $\Delta \gamma^\nu$.

The next step is to introduce an explicit phase-space parametrization for the evaluation of Eqs.~\eqref{eq:S2II},~\eqref{eq:S2CC}. We start with a light-cone (Sudakov) decomposition of the final-state momenta in terms of the two light-like directions along the beam axis $n^\mu$ and $\bar{n}^\mu$, such that $n^2=\bar{n}^2=0$ and $n\cdot \bar{n} = 2$. Any vector $p^\mu$ can be expressed as
\begin{equation}~\label{eq:LC}
    p^\mu = n\cdot p \frac{\bar{n}^\mu}{2}+\bar{n}\cdot p \frac{n^\mu}{2} + p^\mu_\perp=p^+ \frac{\bar{n}^\mu}{2}+p^- \frac{n^\mu}{2} + p^\mu_\perp\,,
\end{equation}
where $p_\perp^2 = - p_t^2 < 0 $. We also introduce the rapidity of $p^\mu$ w.r.t.~the beam axis as
\begin{equation}
    y = \frac{1}{2}\ln\frac{p^-}{p^+}\,.
\end{equation}
Using the decomposition~\eqref{eq:LC}, we can parametrize the phase space in $\Sigma^{(2)}_{\rm II}$ in terms of the phase space of emission $p_1$ and the relative transverse momentum $\zeta \equiv p_{t2}/p_{t1}$, rapidity $\eta \equiv y_2-y_1$ and azimuth $\phi \equiv \phi_2 - \phi_1$ as follows
\begin{equation}\label{eq:param-II}
    [ d p_1 ] [ d p_2] \to  \frac{1}{4}\frac{p_{t1}^2}{(2\pi)^6} d y_1\,d^{2} \vec{p}_{t 1} \,d \eta \,\zeta\,d\zeta \,d\phi\, \exp\left[-\frac{e^{-\gamma_E}}{\nu}2\, p_{t1} \cosh y_1\right]\,,
\end{equation}
where we have neglected the exponential regulator for the second emission $p_2$ since the measurement function \eqref{eq:2L-measure-II} suppresses the integrand for large values of $\eta$.
When expressing the matrix elements ${\cal A}^{(0)}_g(p_1)\,{\cal A}^{(0)}_g(p_2)$ in the $\Sigma^{(2)}_{\rm II}$ integral in terms of this parametrization, we can pull out a factor of $p_{t1}^{-4}$ fixed by dimensional analysis. Since the resulting integrand only depends on $\eta$ and $\phi$ due to invariance under longitudinal boosts and rotations in the transverse plane (in SCET terms, RPI-III \cite{Manohar:2002fd}), we can evaluate the integral over $y_1$ and retain only the term proportional to $\ln\frac{\nu}{p_{t1}}$ as we are interested in the rapidity anomalous dimension. This amounts to the replacement
\begin{equation}
 \int_{-\infty}^{\infty} d y_1 \exp\left[-\frac{e^{-\gamma_E}}{\nu}2\, p_{t1} \cosh y_1\right] = 2\,\ln \frac{\nu}{p_{t1}} + {\cal O}(\nu^0)\,.
\end{equation}
%Checked by RS and SA 28.08.25
%
Finally, given that the product of the amplitudes
${\cal A}^{(0)}_g(p)  \,{\cal A}^{(0)}_g(p)$
in Eq.~\eqref{eq:S2II}  scales as $p_{t1}^{-4}$,
we can perform the $p_{t1}$ integration, which gives rise to 
\begin{align}
    2\int_0^\infty \frac{d p_{t1}}{p_{t1}} &\ln \frac{\nu}{p_{t1}}\,\left(\Theta(p_{T}^{\rm veto} - p_t^{j_1}(p_1,p_2))-\Theta(p_{T}^{\rm veto} - \max\{p_{t1}, p_{t2}\})\right) \notag\\&= 2\,\ln\frac{\nu}{p_{T}^{\rm veto}}\ln\frac{\max\{1,\zeta\}}{\sqrt{1+\zeta^2+2\zeta\cos\phi}}\Theta(\min(d_{1B},d_{2B}) - d_{12}) + {\cal O}(\nu^0)\,,
\end{align}
where we used the measure of Eq.~\eqref{eq:2L-measure-II}. The corresponding contribution to the rapidity anomalous dimension, $\Delta\gamma^{(2)}_{\nu,\,{\rm II}}(R^2)$, is simply given by the coefficient of $\ln\frac{\nu}{p_{T}^{\rm veto}}$.
We discard terms of ${\cal O}(\nu^0)$ which can nevertheless be straightforwardly included with the same approach.
The integrations over $\phi$, $\zeta$ and $\eta$ are well behaved thanks to the measurement function and can be performed numerically.

In the case of $\Sigma^{(2)}_{\rm CC}$, it is more convenient to choose a parametrization that better reflects the structure of collinear singularities in the $\tilde{\cal A}^{(0)}(p_1,p_2)$'s, which diverge when $p_1$ and $p_2$ are collinear. To achieve this, we introduce the parent $k^\mu=p^\mu_1+p^\mu_2$ momentum and set (we recall that for each emission $p^+ p^- = p_t^2$)
\begin{align}\label{eq:parent-vars-1}
    p_1^- = z \, k^-\,&,\quad p_2^- = (1-z) \, k^-\,,\\
    p_{1\perp}^\mu = z \,k_\perp^\mu + z (1-z) q_\perp^\mu\,&,\quad p_{2\perp}^\mu = (1-z) \,k_\perp^\mu - z (1-z) q_\perp^\mu\,,\label{eq:parent-vars-2}
\end{align}
where
\begin{equation}
    q_\perp^\mu = (0,\vec{q}_t,0)\,,\quad k^\mu_\perp = (0,\vec{k}_t,0)\,,
\end{equation}
where $\vec{k}_t$ is the transverse momentum of the parent $k^\mu$. 
We further write $k^- = \sqrt{k_t^2+k^2} e^Y$ with $k^2 = z (1-z) q_t^2$%checked
, and express the phase-space measure as
\begin{equation}\label{eq:param-CC-0}
    [ d p_1 ] [ d p_2] \to \frac{1}{4}\frac{1}{(2\pi)^6} \,d^2 \vec{q}_t \,d Y\, d z\,z\, (1-z)\, d^{2} \vec{k}_t\,   \,\exp\left[-\frac{e^{-\gamma_E}}{\nu}\left(k^++k^-\right)\right]\,.
\end{equation}
where $k_+, k_-$ should be viewed as functions of the integration variables.
This parametrization has the property that it maps the singularities of the $\tilde{\cal A}^{(0)}(p_1,p_2)$ matrix element to the boundaries of the integration region. The configuration where $p_1$ and $p_2$ are soft, in the case of gluons, is mapped to $z\to 0,1$ and the configuration where they are collinear to each other is mapped to $q_t\to 0$. This will prove crucial in the three-loop calculation.
Without loss of generality we can perform a rotation in order to align $\vec{k}_t$ with the $x$ axis in the transverse plane, and write
\begin{equation}\label{eq:qt-kt}
    q_\perp^\mu = q_t \,(0,\cos\theta,\sin\theta,0)\,,\quad k^\mu_\perp = (0,k_t,0,0)\,,\qquad \theta\in [0,2\pi]\,.
\end{equation}
We also introduce $\omega\equiv q_t/k_t$ and recast Eq.~\eqref{eq:param-CC-0} as
\begin{equation}\label{eq:param-CC}
    [ d p_1 ] [ d p_2] \to \frac{1}{4}\frac{k_t^3}{(2\pi)^5} \,\omega\,d \omega \,d\theta\,d Y\, d z\,z\, (1-z)\, d k_t\,   \,\exp\left[-\frac{2e^{-\gamma_E}}{\nu} k_t\,\sqrt{1+\omega^2 z^2 (1-z)^2}\cosh Y\right]\,.
\end{equation}
The generalisation of this parametrization to $D$ dimensions is given in Appendix~\ref{app:details} and will be used for the integration of some of the three-loop contributions.
We can now proceed as in the calculation of $\Sigma^{(2)}_{\rm II}$, and observe that the invariance of the integrand under longitudinal boosts and rotations in the transverse plane guarantees that the matrix element $\tilde{\cal A}^{(0)}(p_1,p_2)$ and the clustering condition are $Y$ independent, which allows us to evaluate the $Y$ integral analytically. Similarly, the $k_t$ integration can be handled as in $\Sigma^{(2)}_{\rm II}$, leading to
\begin{align}
\int d Y\int&\frac{d k_t}{k_t} \,\exp\left[-\frac{2e^{-\gamma_E}}{\nu} k_t\,\sqrt{1+\omega^2 z^2 (1-z)^2}\cosh Y\right]\notag\\
& \times\left(\Theta(p_{T}^{\rm veto} - p_t^{j_1}(p_1,p_2))-\Theta(p_{T}^{\rm veto} - |\vec{p}_{t1} + \vec{p}_{t2}|)\right)\notag\\
%& = -2\,\ln\frac{\nu}{p_{T}^{\rm veto}}\ln\frac{\max\{1,\zeta\}}{\sqrt{1+\zeta^2+2\zeta\cos\phi}}\Theta(d_{12}-\min(d_{1B},d_{2B})) + {\cal O}(\nu^0)\,,\\
& = -2\,\ln\frac{\nu}{p_{T}^{\rm veto}}\ln\frac{\max\{p_{t1}, p_{t2}\}}{|\vec{p}_{t1} + \vec{p}_{t2}|}\Theta(d_{12}-\min(d_{1B},d_{2B})) + {\cal O}(\nu^0)\,,
\end{align}
where the transverse momenta are evaluated with Eqs.~\eqref{eq:parent-vars-2}. The corresponding contribution to the rapidity anomalous dimension, $\Delta\gamma^{(2)}_{\nu,\,{\rm CC}}(R^2)$, is again defined as the coefficient of $\ln\frac{\nu}{p_{T}^{\rm veto}}$. The remaining integrations over $\omega$, $z$ and $\theta$ can be performed numerically with high precision, and we find perfect agreement with the analytic result of Ref.~\cite{Abreu:2022sdc}.

\section{Calculation of $\gamma_{\nu}(p_{T}^{\rm veto},R^2)$ at three loops}
\label{sec:gamma3}
In this section, we outline the structure of the calculation of $\Delta \gamma_{\nu}$ at three loops. Given the technical nature of the three-loop calculation, we report here only the main steps and defer the details to Appendix~\ref{app:details}.

As in the two-loop case, we start from Eq.~\eqref{eq:Dgamma} and consider the $\nu \frac{d}{d\nu}$ derivative of the quantity
\begin{align}\label{eq:start-3L}
    \Sigma^{(3)}&\equiv\Delta {\cal S}^{(3)}(p_{T}^{\rm veto},R^2;\mu,\nu)-\Delta {\cal S}^{(2)}(p_{T}^{\rm veto},R^2;\mu,\nu)\,{\cal S}_{\rm \RSV}^{(1)}(p_{T}^{\rm veto};\mu,\nu)\notag\\
& = \Sigma^{(3)}_{\rm III} + \Sigma^{(3)}_{\rm CCI}+\Sigma^{(3)}_{\rm CCC}\,,
\end{align}
and express the resulting contribution to the rapidity anomalous dimension as
\begin{equation}\label{eq:deltagamma3-dec}
  \Delta\gamma^{(3)}_{\nu}(R^2) =  \Delta\gamma^{(3)}_{\nu,\,{\rm III}}(R^2) + \Delta\gamma^{(3)}_{\nu,\,{\rm CCI}}(R^2) + \Delta\gamma^{(3)}_{\nu,\,{\rm CCC}}(R^2)\,.
\end{equation}
Each of the terms on the r.h.s.~of Eq.~\eqref{eq:start-3L} corresponds to phase-space integrals of squared amplitudes weighted by measurement functions, in a generalisation of Eqs.~\eqref{eq:S2II},~\eqref{eq:S2CC}.
While each term is free of $\epsilon$ poles, making this manifest is not as trivial as at two loops and requires different approaches for the three different types of contributions. 

A new feature that one encounters starting at three loops is that the observable $p_t^{j_1}(p_1,p_2,p_3)$ evaluated on triple-real configurations now depends on the specific choice of jet algorithm, that is on the choice of $p$ in Eq.~\eqref{eq:gen-kt-measures}. As a consequence, different jet algorithms belonging to the generalised-$k_t$ class will have different rapidity anomalous dimensions at three loops and beyond. Due to the complexity of the clustering history, the explicit expression for $p_t^{j_1}(p_1,p_2,p_3)$ in terms of the kinematics of the emissions is quite cumbersome, and we keep it implicit in the equations below. In practice, the observable is evaluated numerically in our computation. An important observation is that, when working in $D$ dimensions, we must use the azimuthal distance defined in Eq.~\eqref{eq:deltaPhiGen}.

We start from the ${\rm III}$ contribution, which arises from the integral of triple independent configurations, that is 
\begin{align}
    \Sigma^{(3)}_{\rm III} &= \frac{1}{3!}\int [d p_1] [d p_2] [d p_3] \,{\cal A}^{(0)}_g(p_1)\,{\cal A}^{(0)}_g(p_2)\,{\cal A}^{(0)}_g(p_3)\notag\\&\times\bigg[\Theta(p_{T}^{\rm veto} - p_t^{j_1}(p_1,p_2,p_3))-\Theta(p_{T}^{\rm veto} - \max\{p_{t1}, p_{t2},p_{t3}\})\notag\\& - \Theta(p_{T}^{\rm veto} - \max\{p_t^{j_1}(p_1,p_2),p_{t3}\})+\Theta(p_{T}^{\rm veto} - \max\{p_{t1}, p_{t2},p_{t3}\})\notag\\& - \Theta(p_{T}^{\rm veto} - \max\{p_t^{j_1}(p_1,p_3),p_{t2}\})+\Theta(p_{T}^{\rm veto} - \max\{p_{t1}, p_{t2},p_{t3}\})\notag\\& - \Theta(p_{T}^{\rm veto} - \max\{p_t^{j_1}(p_2,p_3),p_{t1}\})+\Theta(p_{T}^{\rm veto} - \max\{p_{t1}, p_{t2},p_{t3}\})\bigg]\,.\label{eq:S3III}
\end{align}
The difference of $\Theta$ functions in the second line of the above equation arises from $\Delta {\cal S}^{(3)}$ in Eq.~\eqref{eq:start-3L}, while the differences in the third, fourth and fifth lines arise from the $\Delta {\cal S}^{(2)}\,{\cal S}_{\rm \RSV}^{(1)}$ term. While the integrand of Eq.~\eqref{eq:S3III} does not diverge when two or more of the real particles become collinear to one another, it could still have infrared divergences corresponding to these particles becoming soft. However, thanks to our choice of the RSV observable, these soft divergences cancel locally in Eq.~\eqref{eq:S3III}, so that the phase-space integration can be performed directly in $D=4$.
Due to the absence of singularities when two or more of the emission momenta become collinear, $\Sigma^{(3)}_{\rm III}$ scales as $R^4$ in the small-$R$ limit.
Moreover, by construction this contribution is fully Abelian, and it is proportional to the colour structure $C_R^3$, where we denote by $C_R$ the quadratic Casimir of $SU(N)$ ($C_R=C_F$ for quarks and $C_R=C_A$ for gluons), that is
\begin{equation}
    \Sigma^{(3)}_{\rm III} = C_R^3\,\Sigma^{(3)}_{{\rm III},\,C_R^3}\,.
\end{equation}

We next consider the ${\rm CCI}$ contribution. As for the ${\rm III}$ piece it receives contributions from both the $\Delta {\cal S}^{(3)}$ and the $\Delta {\cal S}^{(2)}\,{\cal S}_{\rm \RSV}^{(1)}$ terms in Eq.~\eqref{eq:start-3L}, and consists of the following two terms
\begin{align}
    \Sigma^{(3)}_{\rm CCI} &= \int [d p_1] [d p_2] [d p_3] \,{\cal A}^{(0)}_g(p_1)\,\left(\frac{1}{2!}\tilde{\cal A}^{(0)}_{gg}(p_2,p_3) +\tilde{\cal A}^{(0)}_{q\bar{q}}(p_2,p_3) \right)\label{eq:S3CCI}\\&\times\bigg[\Theta(p_{T}^{\rm veto} - p_t^{j_1}(p_1,p_2,p_3))-\Theta(p_{T}^{\rm veto} - \max\{p_{t1}, p_t^{j_1}(p_2,p_3)\})\bigg]\notag\\
    & + \int [d p_1] [d p_2] \,{\cal A}^{(0)}_g(p_1){\cal A}^{(1)}_g(p_2)\,\bigg[\Theta(p_{T}^{\rm veto} - p_t^{j_1}(p_1,p_2))-\Theta(p_{T}^{\rm veto} - \max\{p_{t1}, p_{t2}\})\bigg]
    \,.\notag
\end{align}
The first term, in the first two lines of Eq.~\eqref{eq:S3CCI}, arises from configurations with two correlated emissions and a third, independent emission. To write its integral representation, we can choose the correlated emissions to be $p_2$ and $p_3$ in the triple real configuration.\footnote{The ${\rm CCI}$ configuration contributes to the triple-real correction with three permutations, so it receives a combinatorial factor of $1/3$. Here we choose a single one of these permutations and remove the overall factor $1/3$.}
This contribution contains two colour structures and can be parametrized as
\begin{equation}
    \Sigma^{(3)}_{\rm CCI} = C_R^2 C_A\,\Sigma^{(3)}_{{\rm CCI},\,C_R^2C_A}+C_R^2 T_R n_f\,\Sigma^{(3)}_{{\rm CCI},\,C_R^2 T_R n_f}\,.
\end{equation}
Each of these structures is power suppressed in the small-$R$ limits, with leading behaviour $R^2\ln R^2$. This arises from the collinear singularity between the two correlated emissions. 
Unlike for the ${\rm III}$ correction, this triple-real integral diverges in configurations where $p_2$ and $p_3$ are either soft or collinear to each other, giving rise to double poles in $\epsilon$. These divergences are cancelled by the term in the third line of Eq.~\eqref{eq:S3CCI}, which consists of the double-real virtual corrections. To implement this cancellation, the computation of each term in $\Sigma^{(3)}_{\rm CCI}$ must be carried out in $D$ dimensions. 

Finally, the ${\rm CCC}$ contribution arises from $\Delta {\cal S}^{(3)}$ alone and reads
\begin{align}
    \Sigma^{(3)}_{\rm CCC} &= \int [d p_1] [d p_2] [d p_3] \,\left(\frac{1}{3!}\tilde{\cal A}^{(0)}_{ggg}(p_1,p_2,p_3) +\tilde{\cal A}^{(0)}_{gq\bar{q}}(p_1,p^q_2,p^{\bar q}_3) \right)\label{eq:S3CCC}\\&\times\bigg[\Theta(p_{T}^{\rm veto} - p_t^{j_1}(p_1,p_2,p_3))-\Theta(p_{T}^{\rm veto} -  |\vec{p}_{t1}+\vec{p}_{t2}+\vec{p}_{t3}|)\bigg]\notag\\
    & + \int [d p_1] [d p_2] \,\left(\frac{1}{2!}\tilde{\cal A}^{(1)}_{gg}(p_1,p_2) +\tilde{\cal A}^{(1)}_{q\bar{q}}(p_1,p_2) \right)\,\notag\\
    &\times\bigg[\Theta(p_{T}^{\rm veto} - p_t^{j_1}(p_1,p_2))-\Theta(p_{T}^{\rm veto} - |\vec{p}_{t1}+\vec{p}_{t2}|)\bigg]
    \,.\notag
\end{align}
Also in this case, the triple-real integral is not finite, and the corresponding $\epsilon$ poles cancel against the integral of the virtual corrections to the double-soft squared amplitudes.
Therefore, as for $\Sigma^{(3)}_{\rm CCI}$, $\Sigma^{(3)}_{\rm CCC}$ must be evaluated in dimensional regularization. The details of the calculation are discussed in Appendix~\ref{app:details}.
This contribution contains a richer color structure and can be parametrized as
\begin{align}
    \Sigma^{(3)}_{\rm CCC} &= C_R C_F T_R n_f\,\Sigma^{(3)}_{{\rm CCC},\,C_R C_F T_R n_f}+C_R C_A T_R n_f\,\Sigma^{(3)}_{{\rm CCC},\,C_R C_A T_R n_f}\notag\\&+C_R C_A^2\,\Sigma^{(3)}_{{\rm CCC},\,C_R C_A^2}+C_R T_R^2 n_f^2\,\Sigma^{(3)}_{{\rm CCC},\,C_R T_R^2 n_f^2}\,.
\end{align}
Each of these structures scales as $\ln^2 R^2$ in the small-$R$ limit, arising from configurations in which the three correlated particles are collinear with each other.

Following the steps outlined above and Appendix~\ref{app:details}, we have an integral representation for each of the contributions to $\Sigma^{(3)}$. Given the complexity of the clustering algorithm, some integrations are performed numerically. These were done with the \texttt{Cuba} library~\cite{Hahn:2004fe} and cross-checked with \texttt{VegasFlow}~\cite{Carrazza:2020rdn}.

\section{Final results for $\gamma_{\nu}(p_{T}^{\rm veto},R^2)$ and consistency checks}
\label{sec:results}
We now combine Eqs.~\eqref{eq:RAD_RSV} and~\eqref{eq:deltagamma3-dec} to obtain the final result for the three-loop rapidity anomalous dimension. We decompose the anomalous dimension into its colour structures as follows:
\begin{align}\label{eq:gamma3-colour}
    \gamma^{(3)}_{\nu}(R^2) &= C_R C_F T_R n_f\,\gamma_{\nu}^{(3),\,C_R C_F T_R n_f}(R^2)+C_R C_A T_R n_f\,\gamma_{\nu}^{(3),\,C_R C_A T_R n_f}(R^2)\notag\\&+C_R C_A^2\,\gamma_{\nu}^{(3),\,C_R C_A^2}(R^2)+C_R T_R^2 n_f^2\,\gamma_{\nu}^{(3),\,C_R T_R^2 n_f^2}(R^2)\notag\\
    &+C_R^2 C_A\,\gamma_{\nu}^{(3),\,C_R^2C_A}(R^2)+C_R^2 T_R n_f\,\gamma_{\nu}^{(3),\,C_R^2 T_R n_f}(R^2)+C_R^3\,\gamma_{\nu}^{(3),\,C_R^3}(R^2)\,.
\end{align}
As already noted in the previous section, the different colour structures in Eq.~\eqref{eq:gamma3-colour} are also characterised by a different dependence on the jet radius $R$. More specifically, the CCC contributions encoded in the first two lines of Eq.~\eqref{eq:gamma3-colour} scale as $\ln^2 R^2$ for $R\ll 1$. The CCI and III corrections, encoded in the third line of Eq.~\eqref{eq:gamma3-colour}, are power suppressed at small $R$, behaving as $R^2\ln R^2$ and $R^4$ respectively.

To validate our calculation we perform several checks. Firstly, we verify the cancellation of infrared poles in all colour structures. This is non-trivial as in some cases, e.g.~in the $C_R C_A T_R n_f$ colour structure, the cancellation only occurs after combining different contributions from quark and gluon final states. The cancellation of infrared poles is verified numerically, and it provides a stringent consistency check of our setup. 

\begin{figure}
  \centering
\includegraphics[width=0.47\textwidth]{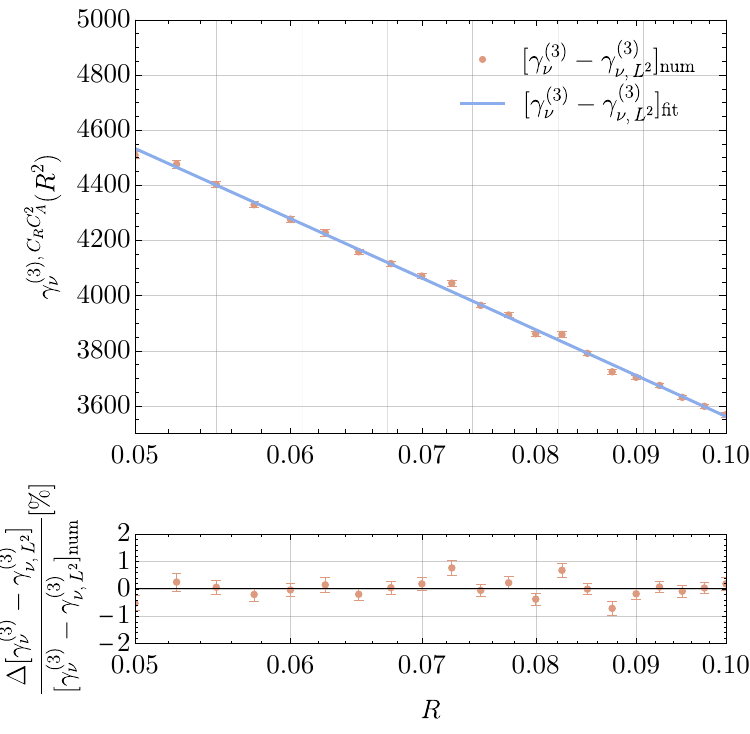} \quad \includegraphics[width=0.487\textwidth]{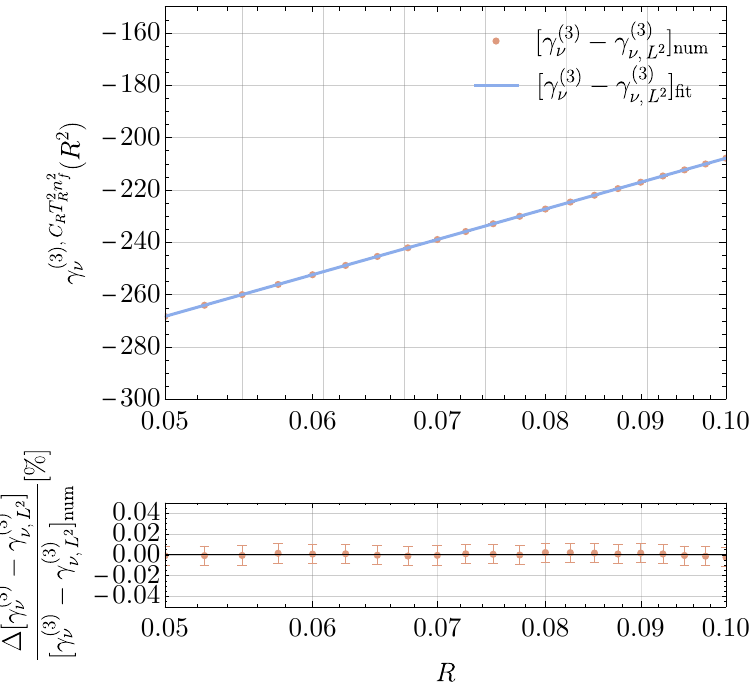} \\
\includegraphics[width=0.47\textwidth]{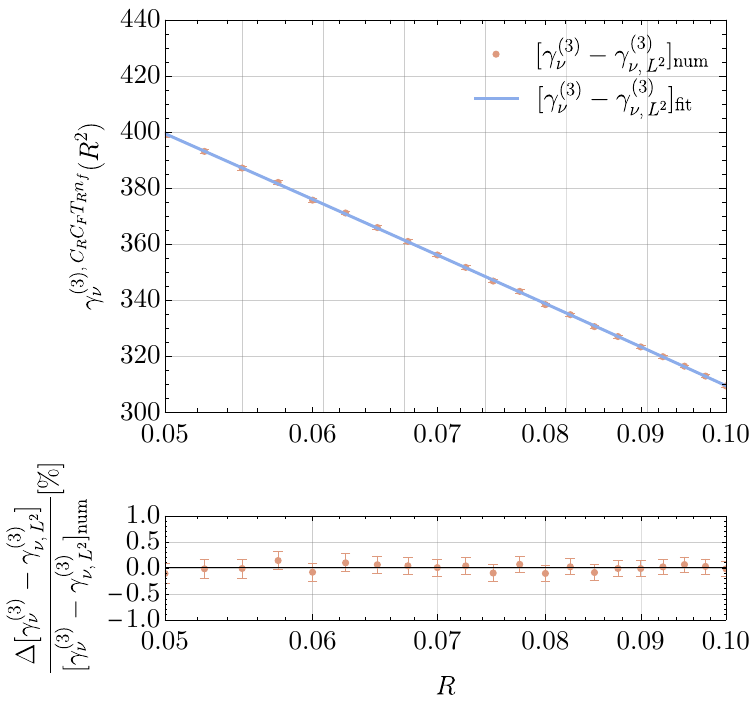} \quad \;\includegraphics[width=0.47\textwidth]{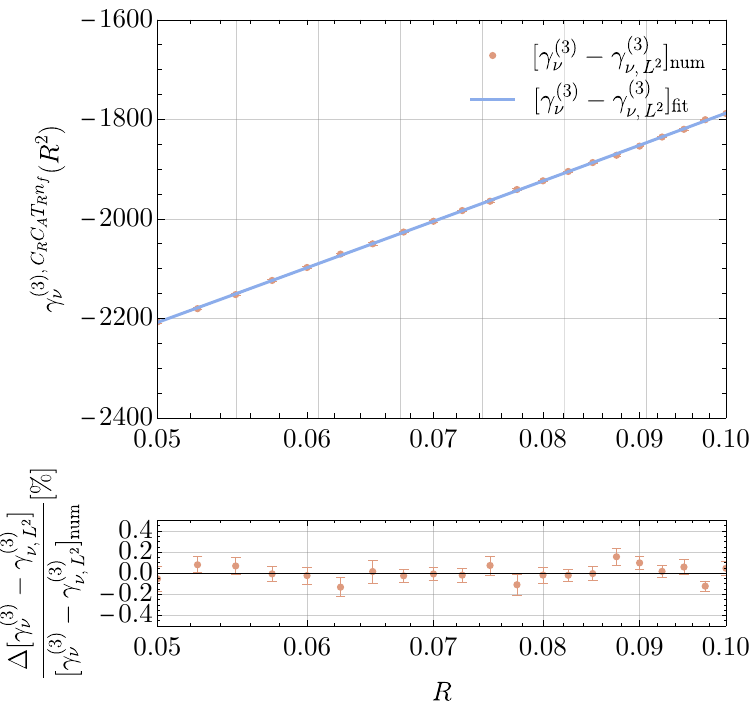}
  \caption{The upper panel of the plots shows, for the four CCC colour structures, the difference between the numerical calculation of the anomalous dimension and the LL prediction ${\cal O}(\ln^2 R^2)$ (points), and a fit with the functional form~\eqref{eq:fit} (solid line). The lower panel shows the difference between the points and the fit. The $x$-axis is logarithmic, to better display the linear dependence in $\ln R^2$ of the difference.}
\label{fig:gamma-smallR}
\end{figure}

As a second check, we verify that we reproduce the known small-$R$ dependence of the anomalous dimension at leading-logarithmic (LL) ${\cal O}(\ln^2 R^2)$ accuracy \cite{Alioli:2013hba,Dasgupta:2014yra,Banfi:2015pju}, namely
\begin{align}\label{eq:smallRLL}
\left.    \gamma^{(3)}_{\nu}(R^2)\right|_{\text{LL small-}R} = 32 \,C_R\,\ln^2 R^2&\left(1.80314\,C_A^2 - 1.17847\,C_A\,T_R\,n_f \right.\notag\\&\left.+ 0.73964\,C_F\,T_R\,n_f - 0.23572\,T_R^2\,n_f^2\right)\,.
\end{align}
This check constrains the CCC contributions, which are the most challenging in our setup.
We find the LL approximation of the anomalous dimension to be rather inaccurate for $R > 0.1$, which demands going to very small $R$ values to perform the check. To carry out the test, we subtract the LL approximation~\eqref{eq:smallRLL} from the full calculation for the four CCC channels (the only ones featuring a logarithmic enhancement), and verify that the resulting function is at most linear in $\ln R^2$. We show this in Fig.~\ref{fig:gamma-smallR}, where the points display the difference between the numerical calculation and the LL approximation, and the blue curve is a fit obtained using the functional form
\begin{equation}\label{eq:fit}
    f_{\rm fit}(R^2) = c_{-1} \, \ln R^2 + c_0 + \sum_{i=1}^{i_{\rm max}} c_i\, (R^2)^{i}\,.
\end{equation}
We perform the fit with different values of $i_{\rm max} \geq 1$, finding stable results under variation of $i_{\rm max}$  (the results in Fig.~\ref{fig:gamma-smallR} are obtained with $i_{\rm max}=1$). The lower panel displays the difference between the points and the fit, showing that the difference is indeed single logarithmic as expected.

\begin{figure}
  \centering
  \includegraphics[width=0.7\textwidth]{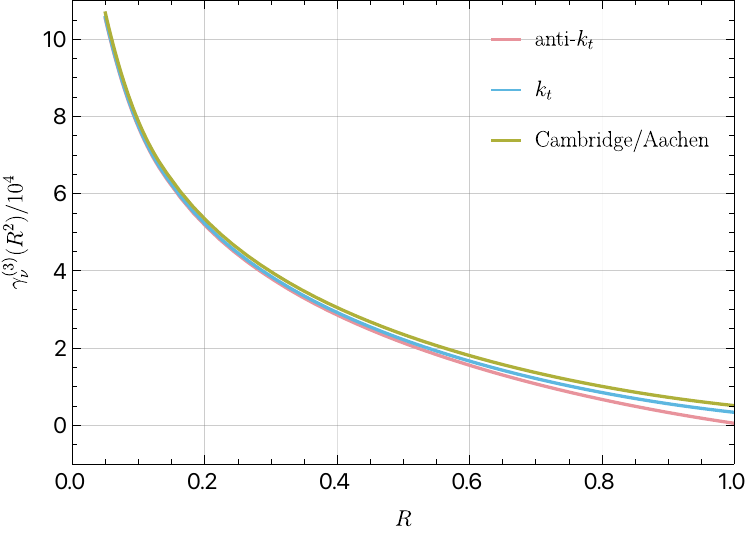}
  \caption{The three-loop rapidity anomalous dimension for jet-veto cross sections (with $C_R=C_A$) in the anti-$k_t$ (red), $k_t$ (green), and Cambridge/Aachen (blue) jet algorithms.}
  \label{fig:gammaNu1}
\end{figure}

\begin{figure}
  \centering
\includegraphics[width=0.463\textwidth]{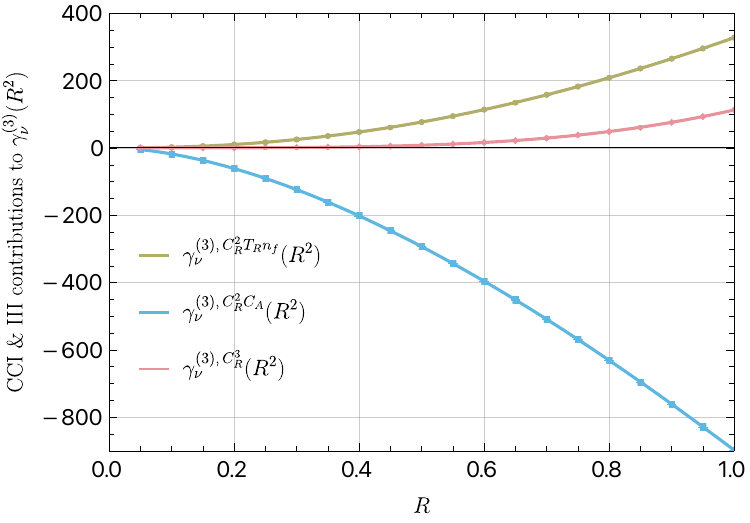} \qquad \includegraphics[width=0.47\textwidth]{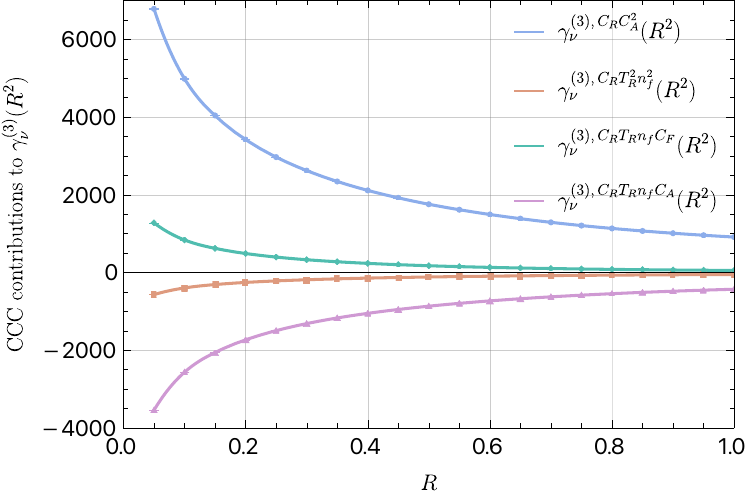} 
  \caption{Breakdown of the three-loop rapidity anomalous dimension for the anti-$k_t$ jet algorithm into the contributing colour structures.}
  \label{fig:gammaNu2}
\end{figure}

The final result for the anomalous dimension~\eqref{eq:gamma3-colour} is reported in Fig.~\ref{fig:gammaNu1} for the three clustering algorithms of the generalised-$k_t$ family ($k_t$, anti-$k_t$, and Cambridge/Aachen), where we set $C_R=C_A$. The corresponding numerical values are also distributed in Ref.~\cite{monni_2026_22914110}.
We recall that this is the first order where the anomalous dimension depends on the jet algorithm, but we note that the variation between the three algorithms we studied is very mild. 
Fig.~\ref{fig:gammaNu2} displays the individual colour structures contributing to the rapidity anomalous dimension in the case of the anti-$k_t$ algorithm. 

\begin{figure}
  \centering
  \includegraphics[width=0.7\textwidth]{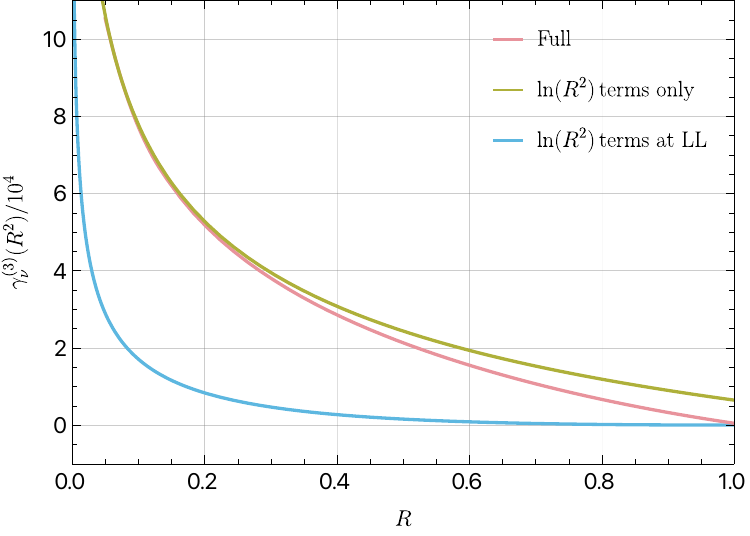}
  \caption{The three-loop rapidity anomalous dimension for the anti-$k_t$ jet algorithm (red), compared to a logarithmic fit (cf.~Eq.~\eqref{eq:smallRNNLL}) at small-$R$  (green) and to the analytic small-$R$ expansion at LL (cyan).}
  \label{fig:gammasmallR}
\end{figure}

It is instructive to compare the full calculation to the result of the small-$R$ fit. Fig.~\ref{fig:gammasmallR} displays a comparison of the full three-loop anomalous dimension (red curve), its small-$R$ approximation at LL~\eqref{eq:smallRLL} (cyan curve), and the full logarithmic fit at small-$R$ (green curve), obtained as follows
\begin{align}\label{eq:smallRNNLL}
\left.    \gamma^{(3)}_{\nu}(R^2)\right|_{\text{small-}R} = \left.    \gamma^{(3)}_{\nu}(R^2)\right|_{\text{LL small-}R} + \left.f_{\rm fit}(R^2)\right|_{c_{i\geq 1}\to 0}\,.
\end{align}
We observe that the LL result is a very poor approximation of the full anomalous dimension, even at very small values of $R$.
On the other hand, the result of the small-$R$ fit is much closer to the exact result, providing an excellent approximation up to values of $R \sim 0.4$, where it overestimates the exact result by $\mathcal O(10\%)$. This result implies that the NLL coefficient is rather large compared to the LL one, making NLL corrections as important as LL ones even at moderately small $R$ values.

To conclude this section, it is useful to report the perturbative series of the rapidity anomalous dimension through ${\cal O}(\alpha_s^3)$ for different values of $R$ (we set $n_f=5$ and $C_R=C_A$):
\begin{align}
    \gamma_\nu(p_{T}^{\rm veto},R^2) = 
    \begin{cases}
  \left(\frac{\alpha_s(p_{T}^{\rm veto})}{4\pi}\right)^2 2174.69+ \left(\frac{\alpha_s(p_{T}^{\rm veto})}{4\pi}\right)^3 77165^{+0.36\permil}_{-0.36\permil} + {\cal O}(\alpha_s^4), & \quad R=0.1, \\
  \left(\frac{\alpha_s(p_{T}^{\rm veto})}{4\pi}\right)^2 1135.60+ \left(\frac{\alpha_s(p_{T}^{\rm veto})}{4\pi}\right)^3 28537^{+0.38\permil}_{-0.38\permil} + {\cal O}(\alpha_s^4), & \quad R=0.4, \\
  \left(\frac{\alpha_s(p_{T}^{\rm veto})}{4\pi}\right)^2 \,\,\,498.59 + \left(\frac{\alpha_s(p_{T}^{\rm veto})}{4\pi}\right)^3 \,\,\,6645^{+0.97\permil}_{-0.97\permil}  + {\cal O}(\alpha_s^4), & \quad R=0.8.
\end{cases}
\end{align}
Working at $p_{T}^{\rm veto}=30$ GeV, with $\alpha_s(p_{T}^{\rm veto}) \simeq 0.142$, we observe a $K$ factor ranging from $K\sim 15\%$--$40\%$ for $R\in [0.1,0.8]$, with the larger values corresponding to smaller jet radii. The impact of this correction on LHC phenomenology will be studied in an upcoming article~\cite{companion}.

\section{Conclusions}
\label{sec:conclusions}
We have completed the calculation of the three-loop rapidity anomalous dimension governing jet-veto cross sections in colour-singlet production. The result is obtained within soft-collinear effective field theory using an exponential rapidity regulator and is presented as a function of the jet radius $R$ for clustering algorithms in the generalised-$k_t$ family. Together with the previously derived two-loop soft function and two-loop beam functions, this provides the last perturbative ingredient required for N$^3$LL resummation of jet-veto logarithms at small veto scales.

The anomalous dimension is extracted from the logarithmic $\nu$ dependence of the three-loop soft function, organised through a decomposition into independent, double-correlated, and triple-correlated contributions. A key aspect of the calculation is the use of an auxiliary reference observable that isolates the $R$ dependence into a finite remainder, enabling a semi-numerical treatment at three loops while preserving analytic control over the renormalisation structure. The results exhibit the expected small-$R$ behaviour, including the logarithmically enhanced terms, and satisfy non-trivial consistency checks such as the cancellation of infrared singularities and agreement with known leading-logarithmic limits.
Within the generalised-$k_t$ family of clustering algorithms, the dependence on the jet algorithm enters for the first time at three loops. We presented results for three standard choices of clustering algorithms ($k_t$, anti-$k_t$ and Cambridge/Aachen) and observe a  mild numerical dependence on the specific choice of algorithm.

With the hard, beam, and soft functions now known to the required fixed order accuracy, the only remaining step towards complete N$^3$LL predictions for jet-vetoed colour-singlet cross sections is the phenomenological implementation and matching to fixed-order results, which we will explore in an upcoming article~\cite{companion}. 
Extensions to related observables, such as jet vetoes in heavy-quark pair production or other SCET$_{\rm II}$ observables with clustering effects, can be pursued using the same strategy developed here.

% \newpage
\section*{Acknowledgments}
We wish to thank Leandro Cieri, Dimitri Colferai, Micha\l{} Czakon, Felix Eschment and Tom Schellenberger for correspondence and cross checks of the implementation of the soft amplitudes computed in Refs.~\cite{Catani:2019nqv,Catani:2022hkb,Czakon:2022dwk},
and Juan Cruz-Martinez for support with the use of \texttt{VegasFlow}.
SA is
supported by the Ram\'on y Cajal program under grant
RYC2024-048348-I.
The work of JRG has been supported by the Royal Society through Grant 
URF\textbackslash{}R1\textbackslash{}201500. The work of RS has been supported by the U.S. Department of Energy under Contract No. DE-SC0012704. The work of PM is funded by the European Union (ERC, grant agreement No. 101044599). Views and opinions expressed are however those of the authors only and do not necessarily reflect those of the European Union or the European Research Council Executive Agency. Neither the European Union nor the granting authority can be held responsible for them.
This research was supported in part by grant NSF PHY-2309135 to the Kavli Institute for Theoretical Physics (KITP).
We thank the Erwin-Schrödinger International Institute for Mathematics and Physics at the University of Vienna for partial support during the Programmes ``Quantum Field Theory at the Frontiers of the Strong Interactions", July 31 - September 1, 2023 and "New Paradigms for Harnessing Quantum Field Theory at Colliders", July 27 - August 28, 2026.

\appendix

\section{Details of the $\Sigma^{(3)}$ calculation}
\label{app:details}

In this appendix, we report some details about the calculation of the $\Sigma^{(3)}$ quantity, discussing in turn the various contributions $\Sigma^{(3)}_{\rm III}$, $\Sigma^{(3)}_{\rm CCI}$ and $\Sigma^{(3)}_{\rm CCC}$.

\subsection{The triple-independent correction $\Delta\gamma^{(3)}_{\nu,\,{\rm III}}(R^2)$}
\label{app:III}
As mentioned in Sec.~\ref{sec:gamma3}, the quantity $\Sigma^{(3)}_{\rm III}$ consists of a finite triple-real phase-space integral that can be calculated in $D=4$ dimensions. 
To compute $\Sigma^{(3)}_{\rm III}$, we proceed as in the $\Sigma^{(2)}_{\rm II}$ two-loop example of Sec.~\ref{sec:deltagamma2}. As noted there, since we are computing a correction to the logarithm of the soft function, this will be at most linear in $\ln\nu/p_{T}^{\rm veto}$. This implies that we have at most a single particle in the triple-real integral that gives rise to a rapidity divergence. Therefore, we can retain the exponential regulator for only one of the three real emissions, say $p_1$. We perform a rotation to align $\vec{p}_{t1}$ with the $x$ axis ($\phi_1=0$), and parametrize the phase space in terms of the variables $\zeta_i\equiv p_{ti}/p_{t1}$, $\phi_i$, and $\eta_i = y_i - y_1$
\begin{equation}\label{eq:param-III}
    [ d p_1 ] [ d p_2] [d p_3]\to \frac{1}{8} \frac{p_{t1}^5}{(2\pi)^8} d y_1\,d p_{t 1} \,d \eta_2 \,d\eta_3 \,\zeta_2\,d\zeta_2\,\zeta_3\,d\zeta_3 \,d\phi_2\,d\phi_3\, \exp\left[-\frac{e^{-\gamma_E}}{\nu}2\, p_{t1} \cosh y_1\right]\,.
\end{equation}
%OK
%
As in the two-loop case, boost invariance ensures that in this parametrization the remaining integrand (matrix element and observable) is independent of $y_1$, and the $p_{t1}$ dependence is fixed by dimensional analysis. We can then integrate over $y_1$ and $p_{t1}$, which amounts to replacing in Eq.~\eqref{eq:S3III} 
\begin{align}
    \int\frac{d p_{t1}}{p_{t1}} &d y_1 \exp\left[-\frac{e^{-\gamma_E}}{\nu}2\, p_{t1} \cosh y_1\right] \\&\times\bigg[\Theta(p_{T}^{\rm veto} - p_t^{j_1}(p_1,p_2,p_3))-\Theta(p_{T}^{\rm veto} - \max\{p_{t1}, p_{t2},p_{t3}\})\notag\\& - \Theta(p_{T}^{\rm veto} - \max\{p_t^{j_1}(p_1,p_2),p_{t3}\})+\Theta(p_{T}^{\rm veto} - \max\{p_{t1}, p_{t2},p_{t3}\})\notag\\& - \Theta(p_{T}^{\rm veto} - \max\{p_t^{j_1}(p_1,p_3),p_{t2}\})+\Theta(p_{T}^{\rm veto} - \max\{p_{t1}, p_{t2},p_{t3}\})\notag\\& - \Theta(p_{T}^{\rm veto} - \max\{p_t^{j_1}(p_2,p_3),p_{t1}\})+\Theta(p_{T}^{\rm veto} - \max\{p_{t1}, p_{t2},p_{t3}\})\bigg]\notag\\&= 
    2\ln\frac{\nu}{p_{T}^{\rm veto}}\ln\frac{\max\{1,p_t^{j_1}(p_2,p_3)\}\max\{\zeta_3,p_t^{j_1}(p_1,p_2)\}\max\{\zeta_2,p_t^{j_1}(p_1,p_3)\}}{\left(\max\{1,\zeta_2,\zeta_3\}\right)^2\,p_t^{j_1}(p_1,p_2,p_3)} + {\cal O}(\nu^0)\,,\notag
\end{align}
where now the observable $p_t^{j_1}(p_1,\dots,p_n)$ introduced in Sec.~\ref{sec:definition} is to be evaluated with the rescaled momenta
\begin{align}
\begin{split}\label{eq:momenta-III}
 p_1 &= p_{t1} (\cosh y_1,\cos\phi_1,\sin\phi_1,\sinh y_1)\,,\\
 p_2 &= \zeta_2 \,p_{t1}(\cosh(y_2-y_1), \cos(\phi_2-\phi_1), \sin(\phi_2-\phi_1), \sinh(y_2-y_1))\,,\\
 p_3 &= \zeta_3 \,p_{t1}(\cosh(y_3-y_1), \cos(\phi_3-\phi_1), \sin(\phi_3-\phi_1), \sinh(y_3-y_1))\,,
\end{split}
\end{align}
where we can set $p_{t1}=1$ and $\phi_1 = y_1 = 0$.
After the above analytic integration we are left with a six-dimensional phase-space integral, that we perform numerically as a function of $R$. More details on the numerical calculation are given in Sec.~\ref{sec:results}.

\subsection{The double-correlated correction $\Delta\gamma^{(3)}_{\nu,\,{\rm CCI}}(R^2)$}
\label{sec:cci}
We now consider $\Sigma^{(3)}_{\rm CCI}$ defined in Eq.~\eqref{eq:S3CCI}. 
An important difference with the $\Sigma^{(3)}_{\rm III}$ correction is that the triple-real and double-real-virtual contributions to $\Sigma^{(3)}_{\rm CCI}$ are individually divergent and must be evaluated in $D$ dimensions. The infrared structure of $\Sigma^{(3)}_{\rm CCI}$ is that of a one-loop calculation, so that the double poles of the double-real-virtual integrand (contained in ${\cal A}^{(0)}_g(p_1){\cal A}^{(1)}_g(p_2)$) are cancelled by the phase space integral of the triple-real squared amplitude.
To manage the cancellation of infrared singularities, we construct a local subtraction scheme tailored to our problem. We add and subtract to Eq.~\eqref{eq:S3CCI} the following local counterterm
\begin{align}
   \Sigma^{\rm (c.t.)}_{\rm CCI}  &= \int [d p_1] [d p_2] [d p_3] \,{\cal A}^{(0)}_g(p_1)\,\left(\frac{1}{2!}\tilde{\cal A}^{(0)}_{gg}(p_2,p_3) +\tilde{\cal A}^{(0)}_{q\bar{q}}(p_2,p_3) \right)\label{eq:ctCCI}\\&\times\bigg[\Theta(p_{T}^{\rm veto} - p_t^{j_1}(p_1,p_2+p_3))-\Theta(p_{T}^{\rm veto} - \max\{p_{t1}, |\vec{p}_{t2}+\vec{p}_{t3}|\})\bigg]\notag\,.
\end{align}
The difference between the triple-real integral in Eq.~\eqref{eq:S3CCI} and the above counterterm is finite and calculable in $D=4$. The integral of $\Sigma^{\rm (c.t.)}_{\rm CCI}$ can be then combined with the double-real-virtual corrections in Eq.~\eqref{eq:S3CCI} resulting in a second finite contribution. We start by handling this second contribution, which can be computed analytically using the techniques of Ref.~\cite{Abreu:2022sdc} as a Laurent expansion in $R^2$. We obtain
\begin{align}\label{eq:cci-ct-R}
    \int [d p_1] [d p_2]& \,{\cal A}^{(0)}_g(p_1){\cal A}^{(1)}_g(p_2)\,\bigg[\Theta(p_{T}^{\rm veto} - p_t^{j_1}(p_1,p_2))-\Theta(p_{T}^{\rm veto} - \max\{p_{t1}, p_{t2}\})\bigg] + \Sigma^{\rm (c.t.)}_{\rm CCI} \notag\\
    & = \frac{\alpha_s^3}{(4\pi)^3}C_R^2 R^2 \ln\frac{\nu}{p_{T}^{\rm veto} }\bigg[8 \left(3 \,R^2-4 \,\pi^2\right)\left(C_A\left(\frac{67}{27}-\frac{\pi^2}{9}\right) - \frac{20}{27} T_R \, n_f\right)\\&+ b_0 \left(-128\, \zeta_3 + 16 R^2 -\frac{4}{9} R^4 -\frac{R^6}{135}-\frac{R^{8}}{3780}+{\cal O}(R^{10})\right)\bigg]\,,\notag
\end{align}
with $b_0$ being defined in Eq.~\eqref{eq:MSbar}. 
We observe that the contribution from Eq.~\eqref{eq:cci-ct-R} is power suppressed in $R^2$, and so is the remaining triple-real integral minus the corresponding local counterterm. We have checked that the $R^2$ expansion agrees well with the exact numerical calculation of the l.h.s. of Eq.~\eqref{eq:cci-ct-R}.

To perform the above calculation, we generalise the phase-space parametrisation of Eq.~\eqref{eq:param-CC-0} to $D$-dimensions, and apply it to the correlated pair $p_2$, $p_3$ in the counterterm integral~\eqref{eq:ctCCI}. We write the light-cone components of the momenta $p_2$ and $p_3$ in the parametrisation of Eq.~\eqref{eq:LC} as
\begin{align}
    p_2^- = z \, k^-\,&,\quad p_3^- = (1-z) \, k^-\,,\\
    p_{2\perp}^\mu = z \,k_\perp^\mu + z (1-z) q_\perp^\mu\,&,\quad p_{3\perp}^\mu = (1-z) \,k_\perp^\mu - z (1-z) q_\perp^\mu\,,
\end{align}
where now $k^\mu=p_2^\mu+p_3^\mu$, $k_\perp^\mu$ is its transverse component, and $q_\perp^\mu$ denotes the auxiliary transverse vector. We can choose these transverse vectors to live in the two-dimensional subspace of the three-body phase space, and we parametrise them according to the $D$-dimensional analogue of Eq.~\eqref{eq:qt-kt}:
\begin{equation}
    q_\perp^\mu = \left(0,\vec{q}_t,0\right)\,,\qquad k_\perp^\mu = \left(0,\vec{k}_t,0\right)\,,
\end{equation}
where
\begin{equation}
    \vec{q}_t = q_t\,\left(\cos\theta,\sin\theta,\vec{0}_{-2\epsilon}\right)\,,\qquad\vec{k}_t = k_t\,\left(1,0,\vec{0}_{-2\epsilon}\right)\,.
\end{equation}
The corresponding phase-space measure becomes
\begin{equation}\label{eq:param-CC-D}
    [ d p_2 ] [ d p_3] \to \frac{1}{4} \,d Y\, d z \,(1-z)^{1-2\epsilon}\, z^{1-2\epsilon}\, \frac{d^{2-2\epsilon} \vec{k}_t\, d ^{2-2\epsilon} \vec{q}_t}{(2\pi)^{6-4\epsilon}}  \,\exp\left[-\frac{e^{-\gamma_E}}{\nu}\left(n\cdot k + \bar{n}\cdot k\right)\right]\,.
\end{equation}

Conversely, the remaining real momentum $p_1^\mu$ now has a generic direction in the $D$-dimensional phase space, that we parametrise by an additional angle $\xi$ describing a fifth dimension as follows
\begin{equation}\label{eq:p1}
  p_1^\mu = p_{t1} \left(\cosh y_1,\cos\phi_1, \sin\phi_1\cos\xi,\sin\phi_1\sin\xi,\vec{0}_{-1-2\epsilon},\sinh y_1\right)\,,
\end{equation}
with $\xi,\phi_1,\theta\in[0,\pi]$. The corresponding phase-space measure is
\begin{equation}\label{eq:dp1}
    [dp_1] = \frac{1}{(2\pi)^{3-2\epsilon}}\frac{\pi^{-\epsilon}}{\Gamma(-\epsilon)} \,d p_{t1} \,p_{t1}^{1-2\epsilon} \,d y_1\,d\phi_1\,\sin^{-2\epsilon}\phi_1\,d\xi\,\sin^{-1-2\epsilon}\xi\,.
\end{equation}

We can now perform the integration over $\vec{q}_t$ and $z$ in the counterterm~\eqref{eq:ctCCI}, and verify that the resulting poles cancel against the infrared divergences of the virtual correction ${\cal A}^{(1)}_g(p_2)$ upon identifying $p^\mu_2$ with $k^\mu$ in the virtual integral~\eqref{eq:cci-ct-R}. One is then left with a finite integral over the components of $p_1$ and $k$. 
However, because of the divergence at $\xi\to 0$, we cannot simply set $\epsilon\to 0$. Instead, we perform a Lorentz rotation to place the remaining momenta $p_1$ and $k$ on the four-dimensional subspace, and perform the $\xi$ integration. The integral over the remaining variables can be evaluated analytically with the method of Ref.~\cite{Abreu:2022sdc} after setting $\epsilon\to 0$.

We are now left with the triple-real integral of Eq.~\eqref{eq:S3CCI} minus the local counterterm~\eqref{eq:ctCCI}. By construction, this integral is locally finite and can be performed in $D=4$ dimensions, for which we adopt the phase-space parametrisation of Eq.~\eqref{eq:param-III}.
As argued previously, only the $y_1$ integration requires a rapidity regulator. Moreover, the invariance of the integrand under longitudinal boosts and rotations in the transverse plane allows us to perform the integration over the rapidity $y_1$ and transverse momentum $p_{t1}$ of the first (independent) emission $p_1$, upon an appropriate rescaling of the kinematics w.r.t.~that of emission $p_1$.
The integration over these variables results in the replacement
\begin{align}
        \int\frac{d p_{t1}}{p_{t1}} &d y_1 \exp\left[-\frac{e^{-\gamma_E}}{\nu}2\, p_{t1} \cosh y_1\right] \\&\times\bigg[\Theta(p_{T}^{\rm veto} - p_t^{j_1}(p_1,p_2,p_3))-\Theta(p_{T}^{\rm veto} - \max\{p_{t1}, p_t^{j_1}(p_2,p_3)\}))\notag\\
        & - \Theta(p_{T}^{\rm veto} - p_t^{j_1}(p_1,p_2+p_3))+\Theta(p_{T}^{\rm veto} - \max\{p_{t1}, |\vec{p}_{t2}+\vec{p}_{t3}|\})\bigg] \notag\\&= 2\ln\frac{\nu}{p_{T}^{\rm veto}}\ln\frac{\max\{1,p_t^{j_1}(p_2,p_3)\}p_t^{j_1}(p_1,p_2+p_3)}{\max\{1,|\vec{p}_{t2}+\vec{p}_{t3}|\}\,p_t^{j_1}(p_1,p_2,p_3)} + {\cal O}(\nu^0)\,,
\end{align}
where the observable $p_t^{j_1}(p_1,\dots,p_n)$ and the total transverse momentum of the correlated pair $|\vec{p}_{t2}+\vec{p}_{t3}|$ are to be evaluated with the rescaled momenta obtained from Eq.~\eqref{eq:momenta-III} by setting $p_{t1} = 1$ and $\phi_1=y_1=0$, as discussed in Sec.~\ref{app:III}.
The remaining six-dimensional integral is well defined and it is performed numerically as a function of $R$.

\subsection{The triple-correlated correction $\Delta\gamma^{(3)}_{\nu,\,{\rm CCC}}(R^2)$}
We finally consider $\Sigma^{(3)}_{\rm CCC}$ defined in Eq.~\eqref{eq:S3CCC}.
As for $\Sigma^{(3)}_{\rm CCI}$, the computation of this term must be performed in $D$ dimensions since it entails a cancellation of infrared singularities between the triple-real (first two lines of Eq.~\eqref{eq:S3CCC}) and double-real-virtual corrections (third and fourth lines of Eq.~\eqref{eq:S3CCC}).
In this case, however, the singularity structure of the triple-real integral is more involved than in the ${\rm CCI}$ case, and it requires a generalisation of the local subtraction performed in Sec.~\ref{sec:cci}.
In particular, the triple-real integrand in Eq.~\eqref{eq:S3CCC} contains divergences when any of the three correlated particles become soft and any subset of them becomes collinear with each other. To deal with this structure, we introduce in the triple-real integral a partition of unity, defined in terms of the following sector functions
\begin{equation}
    \sigma_{ij} = \frac{1}{p_i\cdot p_j}\frac{1}{\sum_{\{k l\}}\frac{1}{p_k\cdot p_l}},\,\quad {\rm with}\,\{ij\},\{k l\} \in \{12,13,23\}\,,
\end{equation}
which satisfy the property
\begin{equation}
    \sum_{\{ij\}} \sigma_{ij} = 1\,.
\end{equation}
The triple-real integral in Eq.~\eqref{eq:S3CCC} then gets decomposed into a sum of three integrals associated with each of the above sectors. Each of these integrals is divergent only when particles $p_i$ and $p_j$ are either soft or collinear to one another. To handle the subtraction of these singularities in each sector, we introduce the following counterterm 
\begin{align}
   {\rm CT}^{\{ij\}}_{\rm CCC}  &= \int [d p_1] [d p_2] [d p_3] \,\left(\frac{1}{3!}\tilde{\cal A}^{(0)}_{ggg}(p_1,p_2,p_3) +\tilde{\cal A}^{(0)}_{gq\bar{q}}(p_1,p^q_2,p^{\bar q}_3) \right)\,\times \sigma_{ij}\label{eq:ctCCC}\\&\times\bigg[\Theta(p_{T}^{\rm veto} - p_t^{j_1}(p_k,p_i+p_j))-\Theta(p_{T}^{\rm veto} -  |\vec{p}_{t1}+\vec{p}_{t2}+\vec{p}_{t3}|)\bigg]\notag\,,
\end{align}
where $p_k$, with $k\neq i,j$ denotes the third momentum in the final state.

When adding and subtracting the three sector counterterms defined as in Eq.~\eqref{eq:ctCCC} to Eq.~\eqref{eq:S3CCC}, we can split the computation of $\Sigma^{(3)}_{\rm CCC}$ into the sum of three triple-real integrals,  three times the integrated counterterm in Eq.~\eqref{eq:ctCCC} and the virtual corrections in Eq.~\eqref{eq:S3CCC}. We will now discuss each of these terms in turn.

The simplest of the three contributions is the integral of the double-real-virtual corrections. Since all the $\epsilon$ poles are already manifest at the integrand level, and no additional singularities arise in the phase-space integration, we can evaluate this integral by performing the $\epsilon$ expansion of the integrand and integrate each term in the expansion numerically. We parametrise the phase space as in Eq.~\eqref{eq:param-CC-D}.
As done in Sec.~\ref{sec:deltagamma2} as well as for the previous corrections to $\Sigma^{(3)}$, we exploit the Lorentz symmetries of the integrand and the presence of a single rapidity divergence to integrate out the transverse momentum $k_t$ and the rapidity $Y$ of the correlated pair (together with the corresponding exponential regulator). This can be done easily by performing the following change of variable
\begin{equation}\label{eq:rescaling}
    q_t = \omega\,k_t\,,
\end{equation}
with $\omega$ being dimensionless. This allows us to factor out, by dimensional analysis, the overall dimension of the integrand in powers of $k_t$. The $Y$  dependence  is, on the other hand, completely fixed by boost invariance~\cite{Manohar:2002fd}, which guarantees that it only appears in the exponential regulator.
These observations allow us to perform the integrations over $Y$ and $k_t$ in $D$ dimensions, which amount to replacing in the double-real-virtual term of Eq.~\eqref{eq:S3CCC}
\begin{align}\label{eq:rapint-v}
\mu^{n\epsilon}\int d Y\int&\frac{d k_t}{k^{1+n\epsilon}_t} \,\exp\left[-\frac{2e^{-\gamma_E}}{\nu} k_t\,\sqrt{1+\omega^2 z^2 (1-z)^2}\cosh Y\right]\\
& \times\left(\Theta(p_{T}^{\rm veto} - p_t^{j_1}(p_1,p_2))-\Theta(p_{T}^{\rm veto} - |\vec{p}_{t1} + \vec{p}_{t2}|)\right)\notag\\
   &= 2\ln\frac{\nu}{p_{T}^{\rm veto}}\left(\frac{\mu}{p_{T}^{\rm veto}}\right)^{n\epsilon}\frac{|\vec{p}_{t1} + \vec{p}_{t2}|^{n\epsilon}-(p_t^{j_1}(p_1,p_2))^{n\epsilon}}{n\epsilon} + {\cal O}(\nu^0)\,,\notag
\end{align}
where $n=6$ for the double-real-virtual amplitude and $n=4$ for the coupling renormalisation term proportional to the tree-level double-real correlated amplitude. All momenta in Eq.~\eqref{eq:rapint-v} are now defined in the rescaled kinematics obtained by applying Eq.~\eqref{eq:rescaling} to the momenta parametrisation and setting $k_{t}=1$. When expanding in $\epsilon$ with the rest of the integrand, we set $\mu = p_{T}^{\rm veto}$ and integrate numerically term by term.

We now turn to the calculation of the counterterm~\eqref{eq:ctCCC}. When evaluating this integral, some care is required since now infrared singularities arise from the phase-space integration. This requires us to work in dimensional regularisation.
Due to the complexity of the triple-real correlated squared amplitudes, we tackle this semi-numerically. For definiteness, and without loss of generality, we will describe the case where $\{i,j\}=\{2,3\}$ and $k=1$. The integration for the remaining two sectors proceeds in an analogous way and produces identical results.
We parametrise the three-body phase space by combining Eqs.~\eqref{eq:param-CC-D},~\eqref{eq:dp1}.
As a first step, we will integrate over the overall scale of the problem and over one of the rapidities to handle the rapidity divergences. This step can be performed via the following changes of variable
\begin{equation}\label{eq:rescaling-ccc}
    q_t^2 = \chi\,p_{t1}^2\,,\quad k_t = \zeta\,p_{t1}\,,\quad Y = y_1 + \Delta Y\,
\end{equation}
in the parametrisation of the momenta. This allows us to integrate out $y_1$ (with the exponential regulator) and the overall scale $p_{t1}$. These integrations simply amount to replacing, in Eq.~\eqref{eq:ctCCC},
\begin{align}\label{eq:rapint-ct}
\mu^{6\epsilon}\int d y_1\int&\frac{d p_{t1}}{p_{t1}^{1+6\epsilon}} \,\exp\left[-\frac{e^{-\gamma_E}}{\nu}2\, p_{t1} \cosh y_1\right]\\&\times\bigg[\Theta(p_{T}^{\rm veto} - p_t^{j_1}(p_1,p_2+p_3))-\Theta(p_{T}^{\rm veto} -  |\vec{p}_{t1}+\vec{p}_{t2}+\vec{p}_{t3}|)\bigg]\notag\\
   &= 2\ln\frac{\nu}{p_{T}^{\rm veto}}\left(\frac{\mu}{p_{T}^{\rm veto}}\right)^{6\epsilon}\frac{|\vec{p}_{t1}+\vec{p}_{t2}+\vec{p}_{t3}|^{6\epsilon}-(p_t^{j_1}(p_1,p_2+p_3))^{6\epsilon}}{6\epsilon} + {\cal O}(\nu^0)\,,\notag
\end{align}
where $\mu$ will be set to $p_{T}^{\rm veto}$. Since we are working in $D$ dimensions, the clustering algorithm used to obtain $p_t^{j_1}(p_1,p_2+p_3)$ is the generalisation of the four-dimensional variant defined in Sec.~\ref{sec:definition}. When performing the $\epsilon$ expansion of the integrand, we must account for the singularities arising from the phase-space integration. In the parametrization of Eqs.~\eqref{eq:param-CC-D},~\eqref{eq:dp1},~\eqref{eq:rescaling-ccc}, the possible singularities arise from $z\to 0,1$ (corresponding to soft divergences in cases where the two correlated partons are gluons), $\chi \to 0$ (corresponding to the two correlated partons in the $\{23\}$ sector becoming collinear), and $\xi\to 0$. The latter is an unphysical singularity but it must be handled consistently in the $\epsilon$ expansion. We then first perform the changes of variable
\begin{equation}
    \chi = \frac{u}{1-u}\,,\quad \cos\xi=2 v -1\,,
\end{equation}
to map all singular integrations onto $[0,1]$, and then isolate the singular contributions with plus distributions by means of the expansion
\begin{equation}
    X^{-1-n\epsilon} = -\frac{1}{n\epsilon} \delta(X) + \left[\frac{1}{X}\right]_+ - n \left[\frac{\ln X}{X}\right]_+ \epsilon + \frac{n^2}{2}\left[\frac{\ln^2 X}{X}\right]_+ \epsilon^2 + \dots\,,
\end{equation}
with $X\in \{z,1-z,u,v,1-v\}$. We can now expand the resulting integrand in $\epsilon$ and integrate the resulting finite expressions numerically, term by term.

Finally, we consider the triple-real correction in Eq.~\eqref{eq:S3CCC}. When combined with (minus) the counterterm, each sector gives rise to a finite integral that we evaluate directly in $D=4$ dimensions. For instance, for sector $\{23\}$, we use the phase-space parametrization in Eqs.~\eqref{eq:param-III},~\eqref{eq:momenta-III} and exploit the symmetries of the problem to integrate out the rapidity $y_1$ and transverse momentum $p_{t1}$. This results in
\begin{align}
        \int\frac{d p_{t1}}{p_{t1}} &d y_1 \exp\left[-\frac{e^{-\gamma_E}}{\nu}2\, p_{t1} \cosh y_1\right] \\&\times\bigg[\Theta(p_{T}^{\rm veto} - p_t^{j_1}(p_1,p_2,p_3))-\Theta(p_{T}^{\rm veto} -  p_t^{j_1}(p_1,p_2+p_3))\bigg]\times\sigma_{\{23\}} \notag\\&= 2\ln\frac{\nu}{p_{T}^{\rm veto}}\ln\frac{p_t^{j_1}(p_1,p_2+p_3)}{p_t^{j_1}(p_1,p_2,p_3)}\times\sigma_{\{23\}} + {\cal O}(\nu^0)\,,
\end{align}
where, as usual, the observable is now evaluated in the rescaled kinematics of Eq.~\eqref{eq:momenta-III}, with $p_{t1}=1$ and $y_1=\phi_1=0$. The remaining two sectors ($\{12\}$, $\{13\}$) can be handled with an analogous procedure.

\bibliographystyle{JHEP} \bibliography{jet}
\end{document}